\documentclass[a4paper,pdftex,reqno,11pt]{amsart}

\allowdisplaybreaks[1]

\usepackage{geometry}

\usepackage{graphicx}
\graphicspath{{Figures/}}

\usepackage{courier}
\usepackage{times}
\usepackage{amsmath,amssymb}
\usepackage[euler-digits]{eulervm}
\usepackage{eurosym}
\usepackage{enumitem}

\usepackage{hyperref}

\newtheorem{Theorem}{Theorem}[section]

\newtheorem{Conjecture}[Theorem]{Conjecture}
\newtheorem{Corollary}[Theorem]{Corollary}
\newtheorem{Lemma}[Theorem]{Lemma}

\newenvironment{Proof}
{\begin{trivlist}\item[]{{\sc Proof.}}}{\hfill{$\square$}\noindent\end{trivlist}}

\newtheorem{Definition}[Theorem]{Definition}

\def\game{v}
\def\voter{voter}

\begin{document}

\title{Measuring voting power in convex policy spaces}
\begin{abstract}
  Classical power index analysis considers the individual's ability to influence the aggregated group decision
  by changing its own vote, where all decisions and votes are assumed to be binary. In many practical applications
  we have more options than either {\lq\lq}yes{\rq\rq} or {\lq\lq}no{\rq\rq}. Here we generalize three important power 
  indices to continuous convex policy spaces. This allows the analysis of a collection of economic problems like
  e.g.\ tax rates or spending that otherwise would not be covered in binary models.
     
  \medskip
     
\end{abstract}
%% 
%% \date{}
%% 
\maketitle

\section{Introduction}

Important decisions are likely made by groups of experts or in the democratic decision-making context by voters.
Giving a set of experts opinions or the votes of the population or representatives within a committee, usually the
aggregated decision is (deterministically) determined according to a certain decision rule. Examples of such decision 
(or aggregation) rules are the majority rule, which selects the alternative that has a majority, or more generally
weighted voting rules like e.g.\ used by the US Electoral College or the EU Council of Ministers.

Given such a decision rule it is quite natural to ask for the individual \textit{power}, by which we understand the 
ability to influence the aggregated decision, of the committee members or voters. A lot of literature is concerned
with measuring this voting power under certain circumstances. But in our opinion the answers given so far are not 
completely satisfactory. To this end we quote \cite{napel2004power}:
\begin{quote}
  {\lq\lq}Scientists who study power in political and economic institutions seem divided into two disjoint
  methodological camps. The first one uses non-cooperative game theory to analyze the impact of explicit decision
  procedures and given preferences over a well-defined -- usually Euclidean -- policy space. The second one stands
  in the tradition of cooperative game theory with much more abstractly defined voting bodies: the considered 
  agents have no particular preferences and form winning coalitions which implement unspecified policies. 
  Individual chances of being part of and influencing a winning coalition are then measured by a \textit{power index}.
  
  \noindent
  \dots
  
  \noindent
  Several authors have concluded that it is time to develop a \textit{unified framework} for measuring decision power
  (cf. \cite{felsenthal2001myths,steunenberg1999strategic}).  
  {\rq\rq}
\end{quote}

A similar opinion is e.g.\ shared in \cite{holler2013reflections,le2012voting}. Within the tradition of \cite{napel2004power}
we try to develop such a unified framework for measuring decision power. Our starting point is the well developed
theory of so-called simple games, see \cite{0943.91005}, within the field of cooperative game theory. The big drawback of
simple games is that both -- the voters and the aggregated decision -- are binary. In \cite{felsenthal1997ternary} the authors
propose \textit{abstention} as a third option for the voters, which they argue to occur quite frequently in practice. 
At the end of their paper they drastically conclude ignoring abstention causes serious errors when evaluating the
power of real-world voting systems:
\begin{quote}
  {\lq\lq}It seems that we are confronted here with a clear-cut case of theory-laden (or theory-biased) observation. Scientists,
  equipped with a ready-made theoretical conception, 'observe' in reality phenomena that fit that conception. And where
  the phenomena do not quite fit the theory, they are at best consciously ignored, but more often actually mis-perceived and
  tweaked into the theoretical mould.{\rq\rq} 
\end{quote}
This statement causes several follow-up papers. In \cite{freixas2003weighted} the authors considered simple games with multiple 
levels of approval. If the voters can choose between $j$ (ordered) alternatives and the aggregated decision is taken 
between $k$ (ordered) alternatives, those games are called $(j,k)$ simple games, so that the examples of
\cite{felsenthal1997ternary} fit in as $(3,2)$-simple games. We remark that other authors consider similar extensions where
the alternatives are not ordered, see e.g.\ \cite{bolger1983banzhaf,bolger1986power,bolger1990characterization,
bolger1993value,bolger2000consistent,bolger2002characterizations}. Power indices for $(j,k)$ simple games were
e.g.\ introduced in \cite{freixas2005banzhaf,freixas2005shapley}, while the basic ideas have been developed for the special
case of $(3,2)$ simple games in earlier papers like e.g.\ \cite{felsenthal1997ternary}.

In this paper we want to consider convex policy spaces with a continuum of alternatives. To keep things simple 
and partially supported by empirical evidence, i.e.\ the authors of \cite{demarzo2003persuasion} show that the
individuals (more-dimensional) opinions can often be well approximated by a $1$-dimensional line, we assume the
policy space to be $1$-dimensional. Moreover we normalize the policy space, of both the input and the output, to
the real interval $[0,1]$. To be more precisely we consider the real interval $[0,1]$ of policy alternatives with
single peaked preferences. In some sense those \textit{games} can be considered as the limit of $(j,k)$ simple games,
where both $j$ and $k$ tend to infinity. Going from binary $\{0,1\}$-decisions to continuous $[0,1]$-decisions allows
the analysis of a collection of economic problems like e.g.\ tax rates or spending that otherwise would not be covered
in binary models.  

Almost the whole literature on voting power is limited to binary or discrete models -- a few exceptions are given by e.g.\ 
\cite{grabisch2011model,maaser2007equal,maaser2012mean,maaser2012note,napel2004power,steunenberg1999strategic}.

The aim of this paper is to propose a generalization of some of the notions for simple games with binary $\{0,1\}$-decisions
to continuous $[0,1]$-decisions in a consistent way. To this end we review some basic definitions and results
for binary simple games in Section~\ref{sec_binary_case}. Some of the usual notation is slightly modified so that 
the coincidence with our definitions for the continuous case becomes more transparent. 
In Section~\ref{sec_j_k} we briefly introduce $(j,k)$ simple games. Directly after, we propose basic definitions
and first results for continuous simple games in Section~\ref{sec_continuous}. In Section~\ref{sec_examples_continuous} 
we give some examples of continuous simple games. Generalized versions of three selected power indices are given in
Section~\ref{sec_power_indices}. The case of vote distributions, where not all alternatives are equipropable, is briefly
treated in Section~\ref{sec_not_equiprobable}. We conclude in Section~\ref{sec_conclusion} and suggest some research
question which may carry forward the development of a unified theory of power measurement. Some power index computations
for examples of certain continuous simple games are delayed to an appendix.  

\section{Binary decision rules}
\label{sec_binary_case}

A binary decision (or aggregation) rule can me modeled as a function $\game:2^N\rightarrow\{0,1\}$ mapping
the coalition $S$ of supporters, i.e.\ those who vote {\lq\lq}yes{\rq\rq}, to the aggregated group decision
$\game(S)\in\{0,1\}$. Within the remaining part of the paper we denote by $N=\{1,2,\dots,n\}$ the set of $n\in\mathbb{N}$
voters and by $2^N$ its power set, i.e.\ the set of its subsets. In the following we specialize those
Boolean functions more and more by requiring desirable properties of a binary decision rule. Quite naturally we require:
\begin{enumerate}
  \item[(1)] if no voter is in favor of a proposal, reject it;
  \item[(2)] if all voters are in favor of a proposal, accept it.
\end{enumerate}  

\begin {Definition}
  \label{def_boolean_game}
  A \textbf{Boolean game} is a function $\game:2^N\rightarrow\{0,1\}$ with $\game(\emptyset)=0$ and $\game(N)=1$. The
  set of all Boolean games on $n$ players is denoted by $\mathcal{B}_n$.
\end{Definition}

We remark that some authors drop the condition $\game(N)=1$. When the group of supporting voters is enlarged we 
would usually expect that the group decision does not change from acceptance to rejection. This is formalized in:

\begin {Definition}
  \label{def_simple}
  A \textbf{simple game} is a Boolean game $\game:2^N\rightarrow\{0,1\}$ such that $\game(S)\le \game(T)$ for all
  $\emptyset\subseteq S\subseteq T\subseteq N$. % and $f(N)=1$. 
  The set of all simple games on $n$ players is denoted by $\mathcal{S}_n$.
\end{Definition}

We call subsets $S\subseteq N$ coalitions. Those coalitions come in two types, i.e.\ either we have $\game(S)=1$
or $\game(S)=0$. We speak of a winning coalition in the first and of a losing coalition in the second case.
The set of all winning or the set of all losing coalitions are both sufficient to uniquely characterize a simple game.
A winning coalition, with the property that all proper subsets are losing, is called minimal winning. Similarly, we
call a losing coalition maximal losing if all of its proper supersets are winning. 

\begin{Definition}
  \label{definition_minimal_winning_and_co}
  Let $\game$ be a Boolean game. By $\mathcal{W}$ we denote the set of all winning and by $\mathcal{W}^m$ we
  denote the set of all minimal winning coalitions of $\game$. Similarly, by $\mathcal{L}$ we denote the set of
  all losing and by $\mathcal{L}^M$ we denote the set of all maximal losing coalitions of $\game$.  
\end{Definition}

We remark that both $\mathcal{W}^m$ and $\mathcal{L}^M$ are sufficient to uniquely characterize a simple game. An
example of a simple game for three players is given by $\mathcal{W}=\Big\{\{1,2\},\{1,2,3\}\Big\}$. The remaining six 
coalitions in $N\backslash\mathcal{W}$ are losing. The unique minimal winning coalition is given by $\{1,2\}$ and
$\mathcal{L}^M=\Big\{\{1,3\},\{2,3\}\Big\}$. 

\begin{Definition}
  \label{def_null_voter}
  A {\voter} that is not contained in any minimal winning coalition is called a \textbf{null} {\voter}.
\end{Definition}  

In our previous example voter~$3$ is a null voter. For the next definition we assume that we let our committee 
decide on a certain proposal and its logical negation. It may be seen as somewhat strange if both proposals
would be accepted under the same preferences of the voters. So anticipating the next definition, we can state that
the most studied simple games are generally proper. 

\begin{Definition}
  \label{def_proper_strong}
  A simple game is called \textbf{proper} if the complement $N\backslash S$ of any winning coalition $S$ is losing. It is called
  \textbf{strong} if the complement $N\backslash S$ of any losing coalition $S$ is winning. A simple game that is both proper and
  strong is called \textbf{constant-sum} (or self-dual, or decisive).
\end{Definition}

The \textit{desirability relation}, introduced in \cite{0083.14301}, assumes a certain intuitive ordering of the {\voter}s:

\begin{Definition}
  \label{def_desirability_relation} Given a simple game, characterized by its set of winning coalitions $\mathcal{W}\subseteq 2^N$, 
  we say that {\voter}~$i\in N$ is \textbf{more desirable}
  as {\voter}~$j\in N$, denoted by $i\succeq j$, if
  \begin{enumerate}
    \item[(1)] for all $S\subseteq N\backslash\{i,j\}$ with $S\cup\{j\}\in\mathcal{W}$, we have $S\cup\{i\}\in\mathcal{W}$;  
    \item[(2)] for all $S\subseteq N\backslash\{i,j\}$ with $S\cup\{i\}\in\mathcal{L}$, we have $S\cup\{j\}\in\mathcal{L}=2^N\backslash\mathcal{W}$.
  \end{enumerate}
  We write $i\simeq j$ if $i\succeq j$,  $j\succeq i$ and use $i\succ j$ as abbreviation for $i\succeq j$, $i\not\simeq j$. 
\end{Definition}

As an abbreviation we write $(\mathcal{W},N)$ for a Boolean game with $\mathcal{W}\subseteq 2^N$ as its set of winning 
coalitions.  

\begin{Definition}
  A simple game $(\mathcal{W},N)$ is called \textbf{complete} if for each pair of {\voter}s $i,j\in N$ we have $i\succeq j$
  or $j\succeq i$. The set of all complete (simple) games on $n$ {\voter}s is denoted by $\mathcal{C}_n$.
\end{Definition}

We remark that our previous example of a simple game is complete and we have $1\simeq 2\succeq 3$.

\begin{Definition}
  Let $(\mathcal{W},N)$ be a complete simple game, where $1\succeq 2\succeq \dots\succeq n$, and $S\subseteq N$ be
  arbitrary. A coalition $T\subseteq N$ is a \textbf{direct left-shift} of $S$ whenever there exists a {\voter} $i\in S$ with
  $i-1\notin S$ such that $T=S\backslash\{i\}\cup\{i-1\}$ for $i>1$ or $T=S\cup\{n\}$ for $n\notin S$. Similarly, a coalition 
  $T\subseteq N$ is a \textbf{direct right-shift} of $S$ whenever there exists a {\voter} $i\in S$ with
  $i+1\notin S$ such that $T=S\backslash\{i\}\cup\{i+1\}$ for $i<n$ or $T=S\backslash\{n\}$ for $n\in S$. A coalition $T$ is a
  \textbf{left-shift} of $S$ if it arises as a sequence of direct left-shifts. Similarly, it is a \textbf{right-shift}
  of $S$ if it arises as a sequence of direct right-shifts. A winning coalition $S$ such that all right-shifts of
  $S$ are losing is called \textbf{shift-minimal} (winning). Similarly, a winning coalition $S$ such that all 
  left-shifts of $S$ are winning is called \textbf{shift-maximal} (losing). By $\mathcal{W}^{sm}$ we denote the
  set of all shift-minimal minimal winning coalitions of $(\mathcal{W},N)$ and by $\mathcal{L}^{sM}$ the set of all
  shift-maximal losing coalitions. 
\end{Definition}

In our example we have $\mathcal{W}^{sm}=\Big\{\{1,2\}\Big\}$ and $\mathcal{L}^{sM}=\Big\{\{1,3\}\Big\}$ since
$\{2,3\}$ is a direct right-shift of $\{1,3\}$. Both of the sets $\mathcal{W}^{sm}$ or $\mathcal{L}^{sM}$ are sufficient 
to uniquely characterize a complete simple game. Some simple games permit a more compact representation using
just $n+1$ non-negative real numbers:

\begin{Definition}
  A simple game $(\mathcal{W},N)$ is \textbf{weighted} if there exists a quota $q>0$ and weights $w_1,\dots,w_n\ge 0$
  such that coalition $S$ is winning if and only if $w(S)=\sum_{i\in S} w_i\ge q$. We denote the corresponding game by
  $[q;w_1,\dots,w_n]$. The set of all weighted (simple) games on $n$ {\voter}s is denoted by $\mathcal{T}_n$.
\end{Definition}

We remark that each weighted game $\game$ admits several weighted representations $[q;w]$, e.g.\ if $[q;w]$ is a
weighted representation for $\game$, then $[\lambda\cdot q;\lambda\cdot w]$ is also a weighted representation for all
$\lambda\in\mathbb{R}_{>0}$. The set of weighted representations of a given weighted game $\game$ is even more involved.
As an example we remark that $[2;1,1,1]$ and $[5;4,3,2]$ represent the same game. Our initial example of a simple
game can be written as $[2;1,1,0]$.

We remark that all weighted simple games are complete. Not every complete simple game is weighted\footnote{We 
have $\mathcal{S}_n\subseteq \mathcal{C}_n\subseteq \mathcal{T}_n$, $\mathcal{S}_n\neq \mathcal{C}_n$ for $n\ge 4$, and 
$\mathcal{C}_n\neq\mathcal{T}_n$ for $n\ge 6$.}, but every simple game
is the intersection of finitely many weighted games. The minimum number needed is called \emph{dimension} of the simple game.
The number $\left|\mathcal{C}_n\right|$ of complete simple games with $n$ voters grows much faster than the number
$\left|\mathcal{T}_n\right|$ of weighted simple games with $n$ voters, see e.g.\ \cite{min_sum_rep}. Similarly, the number
$\left|\mathcal{S}_n\right|$ of simple games with $n$ voters grows much faster than the number $\left|\mathcal{C}_n\right|$
of complete simple games with $n$ voters, see. e.g.\ \cite{kurz2013dedekind}.

In the following we will commonly consider so-called normalized weights $w\in\mathbb{R}_{\ge 0}^n$, where 
$\Vert w\Vert_1=\sum_{i=1}^n w_i=1$. As an abbreviation we use $w(S)=\sum_{i\in S} w_i$ for coalitions $S\subseteq N$. 
For weighted simple games the properties proper, strong, and constant-sum are ultimately linked with the quota $q$:  

\begin{Lemma}
  \label{lemma_characterization_proper}
  A weighted game $\game$ with normalized weights $w\in\mathbb{R}_{\ge 0}^n$, i.e.\ $\Vert w\Vert_1=1$, and quota $q\in(0,1]$
  is proper if and only if there exists a weighted representation with normalized weights $w'\in\mathbb{R}_{\ge 0}^n$ and quota
  $q'\in\Big(\frac{1}{2},1\Big]$.
\end{Lemma}
\begin{Proof}
  If $q>\frac{1}{2}$, then for each winning coalition $S$ with $w(S)\ge q$ we have $w(N\backslash S)=1-w(S)\le 1-q<\frac{1}{2}<q$
  so that $N\backslash S$ has to be losing and the game $\game$ is proper.
  
  For the other direction we assume that $\game$ is proper. 
  From Definition~\ref{def_boolean_game} we conclude $w\neq\mathbf{0}$, so that we can set $w'=\frac{w}{\Vert w\Vert_1}$, 
  i.e.\ we normalize the weights to sum $1$. Next we set
  $$
    q_1=\max_{S\in\mathcal{L}} w'(S)\quad\text{and}\quad
    q_2=\max_{S\in\mathcal{W}} w'(S),
  $$
  where obviously $0\le q_1<q_2\le 1$ due to the definition of a weighted game. Each choice of $q'\in(q_1,q_2]$ 
  corresponds to the same weighted game. Thus it remains to prove that $q_2>\frac{1}{2}$. Assume to the 
  contrary $q_2\le\frac{1}{2}$. Let $S\in\mathcal{W}\neq\emptyset$ be an arbitrary winning coalition. Since
  $$
    w'(N\backslash S)=w'(N)-w'(S)=1-w'(S)\ge 1-q_2\ge \frac{1}{2}\ge q_2,
  $$
  the complementary coalition $N\backslash S$ would be also winning, which is a contraction to the assumption that
  $\game$ is proper.
\end{Proof}

\begin{Lemma}
  \label{lemma_characterization_strong}
  A weighted game $\game$ with normalized weights $w\in\mathbb{R}_{\ge 0}^n$ and quota $q\in(0,1]$ is strong if and
  only if there exists a weighted representation with normalized weights $w'\in\mathbb{R}_{\ge 0}^n$ and quota
  $q'\in\Big(0,\frac{1}{2}\Big]$.
\end{Lemma}
\begin{Proof}
  If $q\le\frac{1}{2}$, then for each losing coalition $S$ with $w(S)< q$ we have $w(N\backslash S)=1-w(S)>1-q\ge \frac{1}{2}\ge q$
  so that $N\backslash S$ has to be winning and the game $\game$ is strong.

  For the other direction we assume that $\game$ is strong. 
  From Definition~\ref{def_boolean_game} we conclude $w\neq\mathbf{0}$, so that we can set $w'=\frac{w}{\Vert w\Vert_1}$, 
  i.e.\ we normalize the weights to sum $1$. Next we set
  $$
    q_1=\max_{S\in\mathcal{L}} w'(S)\quad\text{and}\quad
    q_2=\max_{S\in\mathcal{W}} w'(S),
  $$
  where obviously $0\le q_1<q_2\le 1$ due to the definition of a weighted game. Each choice of $q'\in(q_1,q_2]$ 
  corresponds to the same weighted game. Thus it remains to prove that $q_1<\frac{1}{2}$. Assume to the 
  contrary $q_1\ge \frac{1}{2}$. Let $S\in\mathcal{L}\neq\emptyset$ be an arbitrary losing coalition with $w'(S)=q_1$. Since
  $$
    w'(N\backslash S)=w'(N)-w'(S)=1-w'(S)= 1-q_1\le \frac{1}{2}\le q_1,
  $$
  the complementary coalition $N\backslash S$ would be also losing, which is a contraction to the assumption that
  $\game$ is strong.
\end{Proof}

We remark that given a weighted representation of a weighted game $\game$, the value of the quota $q\in(0,1]$ is not
sufficient to exactly determine whether is proper or non-proper. Similarly $q$ is not sufficient to exactly determine
whether is strong or non-strong. As an example we consider the weighted game $[2;1,1,1]$. For each 
$q\in\Big(\frac{1}{3},\frac{2}{3}\Big]$ the representation $\left[q;\frac{1}{3},\frac{1}{3},\frac{1}{3}\right]$
gives the same game.

\begin{Lemma}
  A weighted game $\game$ is constant-sum if and only if there exists a $\varepsilon>0$ such that for all
  $q\in \left(\frac{1}{2}-\varepsilon,\frac{1}{2}+\varepsilon\right)$ there exists a normalized weighted
  representation with quota $q$.
\end{Lemma}
\begin{Proof}
  If there exists a weighted representation of $\game$ with $q=\frac{1}{2}$, then $\game$ is strong. If there exists
  a weighted representation of $\game$ with $q=\frac{1}{2}+\frac{\varepsilon}{2}>\frac{1}{2}$, then $\game$ is proper.
  
  For the other direction we assume that $\game$ is constant-sum. Let $[q;w]$ we an arbitrary weighted representation of
  $\game$, where $\Vert w\Vert_1=1$. From Definition~\ref{def_boolean_game} we conclude $w\neq\mathbf{0}$, so that we can
  set $w'=\frac{w}{\Vert w\Vert_1}$, i.e.\ we normalize the weights to sum $1$. Next we set
  $$
    q_1=\max_{S\in\mathcal{L}} w'(S)\quad\text{and}\quad
    q_2=\max_{S\in\mathcal{W}} w'(S),
  $$
  where obviously $0\le q_1<q_2\le 1$ due to the definition of a weighted game. Each choice of $q'\in(q_1,q_2]$ 
  corresponds to the same weighted game. From the proofs of Lemma~\ref{lemma_characterization_proper} and 
  Lemma~\ref{lemma_characterization_strong} we conclude $q_2>\frac{1}{2}$ and $q_1<\frac{1}{2}$. This obviously
  admits the choice of a suitable $\varepsilon>0$. 
\end{Proof}

In order to measure the individual's ability to influence the aggregated group decision by changing its own vote, 
a vast amount of so-called power indices was introduced, see e.g.\ \cite{bertini2013comparing}. A common core
is captured by:

\begin{Definition}
  \label{def_power_index}
  Let $\mathcal{V}_n\subseteq \mathcal{B}_n$ a class of Boolean games consisting of $n$~{\voter}s.
  A \textbf{power index} (on $\mathcal{V}_n$) is a mapping $P:\mathcal{V}_n\rightarrow \mathbb{R}^n$.
\end{Definition}

Power indices may have several nice properties:

\begin{Definition}
  \label{def_nice_power_index_properties}
  Let $g:\mathcal{V}_n\rightarrow \mathbb{R}^n=(g_i)_{i\in N}$ be a power index on a class $\mathcal{V}_n$ of
  Boolean games. We say that
  \begin{enumerate}
    \item[(1)] $g$ is \textbf{symmetric}: if for all $\game\in\mathcal{V}_n$ and any bijection $\tau:N\rightarrow N$
               we have $g_{\tau(i)}(\tau \game)=g_i(\game)$, where $\tau \game(S)=\game(\tau(S))$ for any coalition
               $S\subseteq N$;
    \item[(2)] $g$ is \textbf{positive}: if for all $\game\in\mathcal{V}_n$ and all $i\in N$ we have $g_i(\game)\ge 0$
               and $g(\game)\neq 0$;
    \item[(3)] $g$ is \textbf{efficient}: if for all $\game\in\mathcal{V}_n$ we have $\sum_{i=1}^n g_i(\game)=1$;
    \item[(4)] $g$ satisfies the \textbf{null {\voter} property}: if for all $\game\in\mathcal{V}_n$ and all null {\voter}s
               $i$ of $\game$ we have $g_i(\game)=0$.           
  \end{enumerate}
\end{Definition}

Examples of power indices for simple games are e.g.\ the Shapley-Shubik index \cite{shapley1954method}
\begin{equation}
  \operatorname{SSI}_i(\game):= \sum\limits_{S\subseteq N\backslash\{i\}} \frac{|S|!(|N|-1-|S|)!}{|N|!}\cdot
  \left(\game(S\cup\{i\})-\game(S)\right)
\end{equation}
and the absolute Banzhaf index \cite{banzhaf1964weighted}
\begin{equation}
  \operatorname{BZI}_i(\game):= \frac{1}{2^{n-1}}\cdot \sum\limits_{S\subseteq N\backslash\{i\}}\left(\game(S\cup\{i\})-\game(S)\right)
\end{equation}
for all $i\in N$.

We remark that both indices are symmetric, positive and satisfy the null voter property on $\mathcal{B}_n$. The Shapley-Shubik index 
is efficient on $\mathcal{S}_n$, while it is generally not efficient on $\mathcal{B}_n$. Whenever a given positive power index 
$P:\mathcal{V}_n\rightarrow \mathbb{R}^n$ is not efficient, we can consider its \textit{normalization} 
$P_i(\game)/\sum_{j=1}^n P_j(\game)$.

Boolean games where further generalized:

\begin{Definition}
  \label{def_coalitional_game}
  A \textbf{coalitional game} is a function $\game:2^N\rightarrow \mathbb{R}$ with $\game(\emptyset)=0$. 
\end{Definition}  

\begin{Definition}
  \label{def_value}
  Let $\mathcal{V}_n$ be a subclass of coalitional games consisting of $n$~{\voter}s.
  A \textbf{value} (on $\mathcal{V}_n$) is a mapping $P:\mathcal{V}_n\rightarrow \mathbb{R}^n$.
\end{Definition}

Several of the classical power indices have a more general definition as a value. Of course values may
also have some of the properties defined in Definition~\ref{def_nice_power_index_properties}. Additionally
values can \emph{linear}, i.e.\ we have 
\begin{equation}
  P(\lambda_1\cdot\game_1+\lambda_2\cdot\game_2)
  =\lambda_1\cdot P(\game_1)+\lambda_2\cdot P(\game_2)
\end{equation}
for all $\lambda_1,\lambda_2\in\mathbb{R}$ and all coalitional games $\game_1,\game_2\in\mathcal{V}_n$.   

We complete this section with the definition of the third power index, which is studied and generalized in this paper.
To this end let $\game$ be a simple game. We call a vector $x\in\mathbb{R}_{\ge 0}^n$ with $x(N)= 1$ an \textit{imputation}.
The \textit{excess} of a coalition $S$ for imputation $x$ (in $\game$) is given by $e(S,x)=\game(S)-x(S)$. Let $S_1, \dots, S_{2^n}$
be an ordering of all coalitions such that the excess at $x$ is weakly decreasing. The \textit{excess vector} is the
vector $E(x)=(e(x,S_k))_{1\leq k\leq 2^n}$. Imputation $x$ is \textit{lexicographically less} than imputation $y$ if
$E_k(x)<E_k(y)$ for the smallest component $k$ with $E_k(x)\neq E_k(y)$. With this the \textit{nucleolus} is then uniquely
defined as the lexicographically minimal imputation, see e.g.\ \cite{1969}. For the 
weighted game $[3;2,1,1,1]$ the nucleolus is given by $\frac{1}{5}\cdot (2,1,1,1)$, i.e.\ it coincides with the normalized
given weighted representation. The nucleolus has been proposed as a power index e.g.\ in 
\cite{le2012voting,montero2006noncooperative,montero2013nucleolus}.

\section{$\mathbf{(j,k)}$ simple games}
\label{sec_j_k}

In this section we briefly introduce the concept of $(j,k)$ (simple) games, mainly based on \cite{freixas2003weighted}.
So let $j,k\ge 2$ be two arbitrary integers. By $J=\{1,\dots,j\}$ we denote the alternative options in the input and
by $K=\{1,\dots,k\}$ the alternatives in the output. A \textit{numeric evaluation} (of the output, i.e.\ the aggregated
group decision) is a function $\alpha:K\rightarrow\mathbb{R}$ with $\alpha(i)>\alpha(i+1)$ for all $1\le i<k$. Boolean
games, as defined in the previous section, can be seen as $(2,2)$-games, where $J=K=\{1,2\}$ and $\alpha(1)=1$, $\alpha(2)=0$.
In the following we assume the \textit{uniform numeric evaluation} $\alpha(i)=k-i$ and specify all subsequent definitions
without $\alpha$.

\begin{Definition}
  A sequence $S=(S_1,\dots,S_j)$ of mutually disjoint sets $S_i\subseteq N$ with
  $\cup_{1\le i\le j} S_i$ is called \textbf{ordered $\mathbf{j}$-partition}. 
\end{Definition}

For the binary case $j=2$, the set $S_1$ is given by the {\lq\lq}yes{\rq\rq}-voters and $S_2$ by the {\lq\lq}no{\rq\rq}-voters. 
By $J^N$ we denote the set of all ordered $j$-partitions of $N$, i.e.\ we especially have $\left|J^N\right|=j^n$. An ordered
$j$-partition $S=(S_1,\dots,S_j)$ can also be written as a mapping $\beta:N\rightarrow J$, where $S_h=\{i\in N\,:\, \beta(i)=h\}$ 
for all $1\le h\le j$.

\begin{Definition}
  For two ordered $j$-partitions $S=(S_1,\dots,S_j)$, $T=(T_1,\dots,T_j)$ we write $S\overset{j}{\subseteq} T$ if
  $$
    \cup_{1\le h\le i} S_i\subseteq \cup_{1\le h\le i} T_i 
  $$
  for all $1\le i\le j$.
\end{Definition}

\begin{Definition} 
  \label{def_boolean_j_k}
  A \textbf{Boolean $\mathbf{(j,k)}$ game} is given by a function $\game:J^N\rightarrow K$ with $\game\Big((\emptyset,\dots,\emptyset,N)\Big)=k$
  and $\game((N,\emptyset,\dots,\emptyset)\Big)=1$.
\end{Definition}

In other words Definition~\ref{def_boolean_j_k} says that if all voters are in favor of the \textit{lowest} alternative, then
the aggregated group decision should be the \textit{lowest} alternative and similarly for the \textit{highest} alternative. 

\begin {Definition}
  \label{def_simple_j_k}
  A \textbf{$\mathbf{(j,k)}$ simple game} is a Boolean game $\game:J^N\rightarrow K$ such that $\game(S)\le \game(T)$ for all
  ordered $j$-partitions $S\overset{j}{\subseteq} T$. 
\end{Definition}

Let us consider the following example (taken from \cite{felsenthal2001models}) of a $(3,2)$ simple game $\game$ given by 
\begin{eqnarray*}
  \game\big(S_1,S_2,S_3\big) &=& 2\quad\forall S_1\subseteq N\backslash\{1\},\\
  \game\big(S_1,S_2,S_3\big) &=& 2\quad\forall \{2,3\}\subseteq S_3\subseteq N,\\
  \game\big(S_1,S_2,S_3\big) &=& 1\quad\forall \{1\}\subseteq S_1\subseteq N, \left|S_3\cap \{2,3\}\right|\le 1,
\end{eqnarray*}
where $S_1,S_2,S_3$ form an ordered $3$-partition of $N=\{1,2,3\}$. In other words the aggregated group decision is $1$ if and only 
if voter $1$ is in favor of alternative~$1$ and not both of the remaining voters are in favor of alternative~$3$.   

We remark that there are also notions for \textit{complete} or \textit{weighted} $(j,k)$ games, but according to 
\cite{freixas2005shapley} no completely satisfactory definition of a weighted $(j,k)$ game has been found so far, while
several suggestions have been proposed in the literature.

In order to state the Shapley-Shubik and the Banzhaf index for $(j,k)$ simple games, see \cite{freixas2005shapley} and
\cite{freixas2005banzhaf}, we need a few further definitions.

\begin{Definition} 
  A \textbf{queue} of $N=\{1,\dots,n\}$ is a bijection from $N$ to $N$. The set of all queues is denoted by
  $\mathcal{Q}_n$, i.e.\ $\left|\mathcal{Q}_n\right|=n!$.
\end{Definition}

\begin{Definition}
  \label{def_i_pivot}
  Let $\game:J^N\rightarrow K$ be a $(j,k)$ simple game, $q\in\mathcal{Q}_n$ a queue, and $S=(S_1,\dots,S_j)$ an
  ordered $j$-partition. For each $1\le i\le k-1$ the \textbf{$\mathbf{i}$-pivot} is uniquely defined
  either as
  \begin{enumerate}
    \item[(1)] the voter, whose vote in $S$ clinches the aggregated group decision under, at least the output level $i$, 
               independently of the subsequent voters of $i$ in $q$, or
    \item[(2)] the voter, whose vote in $S$ clinches the aggregated group decision under, at most the output level $i+1$, 
               independently of the subsequent voters of $i$ in $q$.             
  \end{enumerate}
\end{Definition}

For the example of the $(3,2)$ game $\game$ above, we consider the queue $q=(2,1,3)$ and the ordered $3$-partition 
$S=\left(\{1\},\{2\},\{3\}\right)$. Since $\game\big(\{1,3\},\{2\},\emptyset\big)=1$ and 
$\game\big(\emptyset,\{2\},\{1,3\}\big)=2$, voter~$2$ is not a $1$-pivot for $q$ and $S$ in $\game$. Since
$\game\big(\{1\},\{2\},\{3\}\big)=\game\big(\{1\},\{2,3\},\emptyset\big)=\game\big(\{1,3\},\{2\},\emptyset\big)=1$, 
voter~$1$ is not a $1$-pivot for $q$ and $S$ in $\game$ and voter~$3$ is a $1$-pivot for $q$ and $S$ in $\game$.

\begin{Definition}
  The Shapley-Shubik index of a $(j,k)$ simple game is given by
  \begin{equation}
    \operatorname{SSI}_i(\game)=\frac{1}{n!\cdot j^n}\cdot\sum_{h=1}^{k-1} \left|\left\{(q,S)\in\mathcal{Q}_n\times J^N
    \,:\,i\text{ is a }h\text{-pivot for $q$ and $S$ in $\game$}\right\}\right|
  \end{equation}
  for all $i\in N$.
\end{Definition}  

For the stated example of the $(3,2)$ simple game we obtain the following pivot-counts per permutation:
\begin{eqnarray*}
  (1,2,3) \rightarrow (18,6,3) \quad\quad (2,1,3) \rightarrow (24,0,3) \quad\quad (3,1,2) \rightarrow (24,3,0) \\
  (1,3,2) \rightarrow (18,3,6) \quad\quad (2,3,1) \rightarrow (24,0,3) \quad\quad (3,2,1) \rightarrow (24,3,0)
\end{eqnarray*}
so that $\operatorname{SSI}(\game)=\left(\frac{22}{27},\frac{5}{54},\frac{5}{54}\right)$.

\begin{Definition}
  Given a $(j,k)$ simple game $\game$ and an ordered $j$-partition $S\in J^N$, we denote by $S_{i\downarrow}$
  the unique ordered $j$ partition which satisfies
  \begin{enumerate}
    \item[(1)] $S_h\backslash\{i\}=T_h\backslash\{i\}$ for all $1\le h\le j$ and
    \item[(2)] $i \in T_{\max(h+1,j)}$ for the index $h$ with $i\in S_h$,
  \end{enumerate}
  where we use the abbreviation $T=S_{i\downarrow}$. The pair $(S,S_{i\downarrow})\in J^N\times J^N$
  is called an \textbf{$\mathbf{(m,l)}$-swing for voter $\mathbf{i}$} if $1\le l<m\le k$, $\game(S)=l$, 
  and $\game(T)=m$. The number of all $(m,l)$-swings for voter~$i$ in $\game$ is denoted by $\eta_i(\game)$.
\end{Definition}
      
\begin{Definition}
  The absolute Banzhaf index of a $(j,k)$ simple game is given by
  \begin{equation}
    \operatorname{BZI}_i(\game)=\frac{1}{j^{n-1}\cdot (k-1)}\cdot\eta_{i}(\game)
  \end{equation}
  for all $i\in N$.
\end{Definition}

We remark that the normalization factor $\frac{1}{j^{n-1}\cdot (k-1)}$, which is not contained in the definition stated in
\cite{freixas2005banzhaf}, is rather debatable, but this way our definitions coincide with the usual definitions for
$(2,2)$ simple games. In most applications the absolute Banzhaf index is normalized to be efficient anyway.

For the stated example of the $(3,2)$ simple game we obtain $\eta_1(\game)=8$, $\eta_2(\game)=\eta_3(\game)=1$, 
so that $\operatorname{BZI}=\frac{1}{9}\cdot\left(8,1,1\right)$ and that the normalized Banzhaf index is given by 
$\left(\frac{4}{5},\frac{1}{10},\frac{1}{10}\right)$. We remark that for this specific example the $\Vert\cdot\Vert_1$-norm
of the difference of the Shapley-Shubik index and the normalized Banzhaf index is given by $\frac{4}{135}\approx 3\%$.

To the best of our knowledge the nucleolus has not been defined for $(j,k)$ simple games so far. In 
Section~\ref{sec_power_indices} we will extend the definition of the nucleolus of simple games to continuous
decision rules. The underlying, rather natural idea, can be used to define the nucleolus also for $(j,k)$ simple games.

For the special case of $(3,2)$ simple games some more power indices were defined in \cite{freixas2012probabilistic}.

\section{Continuous decision rules}
\label{sec_continuous}

In order to rewrite the definitions and results of Section~\ref{sec_binary_case} for the continuous
interval $[0,1]$ instead of the binary set $\{0,1\}$ of alternatives, we identify $2^N$ with $\{0,1\}^n$,
i.e.\ subsets are mapped to incidence vectors
$$
  S\subseteq N \mapsto \left(x_1,\dots,x_n\right)\in\{0,1\}^n,
$$
where $x_i=1$ if $i\in S$ and $x_i=0$ otherwise. An example is given by the incidence vector $(1,0,1,0)$ for
the coalition $\{1,3\}\subseteq\{1,2,3,4\}$.

\begin {Definition}
  \label{def_cont_boolean_game}
  A \textbf{continuous Boolean game} is a function $\game:[0,1]^n\rightarrow[0,1]$ with $\game\big((0,\dots,0)\big)=0$
  and $\game\big((1,\dots,1)\big)=1$. The set of all continuous Boolean games on $n$ players is denoted by $\mathbb{B}_n$.
\end{Definition}

We remark that requiring $\game\big((x,\dots,x)\big)=x$ for all $x\in[0,1]$ would be a rather strong condition, which
is violated by several of the examples considered later on.

Each Boolean game $\game$ can be embedded in a continuous Boolean game $\game'$ extending the function defined
on the $2^n$ points $\{0,1\}^n$ in an arbitrary way. If $\game$ is linear, i.e.\ if $\game(S\cup T)=\game(S)+\game(T)$
for all $S,T\subseteq N$ with $S\cap T$, then we can use convex combinations of the $2^n$ points $\{0,1\}^n$ to
interpolate the intermediate points in $[0,1]^n$. Another way is to use a partition $S,T$ of $[0,1]$ in the following way:
For each $x'\in[0,1]^n$ we define $x\in\{0,1\}^n$ by $x_i=0$ if $x_i'\in S$ and $x_i=1$ if $x_i'\in T$. With this we can
set $\game'(x')=\game(x)\in\{0,1\}$. We remark that using this \textit{threshold-type} embedding the definitions stated
in this section are transfered back to those from Section~\ref{sec_binary_case}.    

Instead of winning and losing coalitions we can look more generally at $z$-coalitions $\{i\in N\mid x_i=z\}$,
$\underline{z}$-coalitions $\{i\in N\mid x_i\le z\}$, and $\overline{z}$-coalitions $\{i\in N\mid x_i\ge z\}$
for each $z\in[0,1]$. So winning coalitions are $1$-coalitions and losing coalitions are $0$-coalitions if
all $x_i$ are binary.

\begin{Definition}
  For two vectors $x=(x_1,\dots,x_n)\in \mathbb{R}^n$ and $y=(y_1,\dots,y_n)\in\mathbb{R}^n$
  we write $x\le y$ if $x_i\le y_i$ for all $1\le i\le n$.
\end{Definition}

\begin {Definition}
  \label{def_cont_simple}
  A \textbf{continuous simple game} is a continuous Boolean game $\game:[0,1]^n\rightarrow[0,1]$ such that 
  $\game(S)\le \game(T)$ for all real-valued vectors $\mathbf{0}\le S\le T\le \mathbf{1}$, where $\mathbf{0}$ denotes 
  the all-$0$- and $\mathbf{1}$ the all-$1$-vector. The set of all continuous simple games on $n$ players is denoted
  by $\mathbb{S}_n$.
\end{Definition}

\begin{Definition}
  \label{def_cont_null_voter}
  Given a continuous Boolean $\game:[0,1]^n\rightarrow[0,1]$, each {\voter} $i\in N$ such that
  $$\game\big((x_1,\dots,x_n)\big)=\game\big((x_1,x_{i-1},x_i',x_{i+1,}\dots,x_n)\big)$$ for all
  $x_1,\dots,x_n,x_i'\in[0,1]$ is called a \textbf{null} {\voter}.
\end{Definition}  

We remark that Definition~\ref{def_cont_null_voter} is equivalent to Definition~\ref{def_null_voter} if all $x_i$ are
binary.

\begin{Definition}
  \label{def_cont_proper_strong}
  A continuous simple game $\game:[0,1]^n\rightarrow[0,1]$ is called \textbf{proper} if $\game(x)+\game(\mathbf{1}-x)\le 1$
  for all real-valued vectors $x\in[0,1]^n$. It is called \textbf{strong} if $\game(x)+\game(\mathbf{1}-x)\ge 1$.
  A continuous simple game that is both proper and strong is called \textbf{constant-sum} (or self-dual, or decisive).
\end{Definition}

In analogy to Definition~\ref{def_desirability_relation} we generalize Isbell's desirability relation as follows:

\begin{Definition}
  \label{def_const_desirability_relation} Given a continuous simple game $\game:[0,1]^n\rightarrow[0,1]$ we say that
  {\voter}~$i\in N$ is \emph{more desirable}
  as {\voter}~$j\in N$, denoted by $i\succeq j$, if
  \begin{enumerate}
    \item[(1)] $\game(\tau(x))\ge \game(x)$ for all $x\in[0,1]^n$ with $x_i\le x_j$, where $\tau$ is equal to the
                transposition $(i,j)$;  
    \item[(2)] $\game(\tau(x))\le \game(x)$ for all $x\in[0,1]^n$ with $x_i\ge x_j$, where $\tau$ is equal to the
                transposition $(i,j)$.
  \end{enumerate}
  We write $i\simeq j$ if $i\succeq j$,  $j\succeq i$ and use $i\succ j$ as abbreviation for $i\succeq j$, $i\not\simeq j$. 
\end{Definition}

We can easily check that Definition~\ref{def_const_desirability_relation} is equivalent to Definition~\ref{def_desirability_relation}
if all $x_i$ are binary.

\begin{Definition}
  A continuous simple game $\game:[0,1]^n\rightarrow[0,1]$ is called \textbf{complete} if for each pair of {\voter}s $i,j\in N$
  we have $i\succeq j$ or $j\succeq i$. The set of all continuous complete (simple) games on $n$ {\voter}s is denoted
  by $\mathbb{C}_n$.
\end{Definition}

As mentioned in the previous section, no completely satisfactory definition of a weighted $(j,k)$ game has been found so far, 
so that we propose several versions of weightedness in the case of continuous simple games. 

\begin{Definition}
  A continuous simple game $\game:[0,1]^n\rightarrow[0,1]$ is \textbf{linearly weighted} if there exist (normalized) weights
  $w_1,\dots,w_n\ge 0$ with $\sum_{i=1}^nw_i=1$ such that $\game\big((x_1,\dots,x_n)\big)=\sum_{i=1}^n w_ix_i$.  
  The set of all continuous linearly weighted (simple) games on $n$ {\voter}s is denoted by $\mathbb{L}_n$.
\end{Definition}

\begin{Definition}
  A continuous simple game $\game:[0,1]^n\rightarrow[0,1]$ is called a \textbf{threshold} game if there exists a quota $q\in(0,1]$
  and (normalized) weights $w_1,\dots,w_n\ge 0$ with $\sum_{i=1}^nw_i=1$ such that $\game\big((x_1,\dots,x_n)\big)=1$
  if $\sum_{i=1}^n w_ix_i\ge q$ and $\game\big((x_1,\dots,x_n)\big)=0$ otherwise.  
  The set of all continuous threshold games on $n$ {\voter}s is denoted by $\mathbb{T}_n$.
\end{Definition}

\begin{Definition}
  A continuous simple game $\game:[0,1]^n\rightarrow[0,1]$ is \textbf{weighted} if there exist (normalized) weights
  $w_1,\dots,w_n\ge 0$ with $\sum_{i=1}^nw_i=1$ and a monotonously  increasing quota function $q:[0,1]\rightarrow[0,1]$ 
  such that $\game\big((x_1,\dots,x_n)\big)=q\left(\sum_{i=1}^n w_ix_i\right)$.  
  The set of all continuous weighted (simple) games on $n$ {\voter}s is denoted by $\mathbb{W}_n$.
\end{Definition}

The quota function $q$ of a weighted continuous simple game satisfies $q(0)=0$ and $q(1)=1$.  
We remark that all linearly weighted, all threshold, and all weighted continuous simple games are complete. 

\begin{Lemma}
  The weighted representation of a continuous linearly weighted game is unique. 
\end{Lemma}
\begin{Proof}
  Assume that a given continuous linearly weighted game $\game$ has two weighted representations $w$ and $\hat{w}$ with
  $w,\hat{w}\in[0,1]^n$ and $\Vert w\Vert_1=\Vert\hat{w}\Vert_1=1$. Thus we have
  $
    \game(x)=w^Tx=\hat{w}^Tx
  $ 
  for all $x\in[0,1]^n$. Inserting $x=(0,\dots,0,1,0,\dots,0)$, where the $1$ is at position $1\le i\le n$, 
  yields $w_i=\hat{w}_i$, so that $w=\hat{w}$.
\end{Proof}

\begin{Lemma}
  Let $\game$ be a continuous threshold game such that there exists a vector $x\neq\mathbf{1}$
  with $\game(x)=1$ and $x_j<1$ for a non-null {\voter} $j$. The weighted representation of $\game$, consisting
  of a quota $q\in(0,1]$ and weights $w\in[0,1]^n$ with $\Vert w\Vert_1=1$, is unique.
\end{Lemma}
\begin{Proof}
  Let $(q^{(1)},w^{(1)})$ and $(q^{(2)},w^{(2)})$ be two representations of $\game$, i.e.\
  $$
    \sum_{i=1}^n w_i^{(1)}\tilde{x}_i\ge q^{(1)} \quad\Longleftrightarrow\quad
    \sum_{i=1}^n w_i^{(2)}\tilde{x}_i\ge q^{(2)}
  $$
  for all $\tilde{x}\in[0,1]^n$. Since $q^{(1)}\in[0,1]$ there exists a $\hat{x}\in[0,1]^n$ with $\hat{x}^Tw^{(1)}=q^{(1)}$.
  Each voter $1\le i\le n$ with $w_i^{(1)}=0$ or $w_i^{(2)}=0$ is a null {\voter}.

  Assume $q^{(1)}=1$: For $x\neq\mathbf{1}$ there is a non-null {\voter} $1\le j\le n$ with $x_j<1$. Thus $w_j^{(1)}=0$ since
  otherwise $x^Tw^{(1)}<1$, which is contradictory to $j$ being a non-null {\voter}. Thus we have $q^{(1)}\neq 1$.  
  
  Next we show that each null {\voter} has zero weight. Let $e_j$ denote the $j$-th unit vector and let $i$ be a
  null {\voter}. If $\hat{x}_i>0$ then $w^{(1)}(\hat{x}-\varepsilon\cdot e_i)<q^{(1)}$ for all $\varepsilon>0$, so
  that $\game(\hat{x}-\varepsilon\cdot e_i)=0$ (for suitably small $\varepsilon$), which is contradictory to $i$ being a
  null {\voter}. If otherwise $\hat{x}_i=0$ for all null {\voter}s $i$, then there exists a non-null {\voter} $j$
  with $\hat{x}_j>0$ and $w_j^{(1)}>0$. With this we have $w^{(i)}(\hat{x}-\varepsilon\cdot e_j)<q^{(1)}$ for
  $\varepsilon>0$ and $\game(\hat{x}-\varepsilon\cdot e_j)=0$ (for suitably small $\varepsilon$). Since
  $w^{(1)}(\hat{x}-\varepsilon\cdot e_j+\varepsilon\cdot \frac{w_j^{(1)}}{w_i^{(1)}}\cdot e_i)=q^{(1)}$, where
  $i$ is an arbitrary null {\voter}, we have $\game(\hat{x}-\varepsilon\cdot e_j+\varepsilon\cdot \frac{w_j^{(1)}}{w_i^{(1)}}\cdot e_i)=1$,
  which contradicts the fact that $i$ is a null {\voter}. Thus each null {\voter} has zero weight.   
  
  Due to symmetry we can state $q^{(1)},q^{(2)}\in (0,1)$, $w_j^{(1)},w_j^{(2)}\in (0,1]$ for all non-null {\voter}s
  $j$, and $w_i^{(1)}=w_i^{(2)}=0$ for all null {\voter}s $i$. If there exists exactly one non-null {\voter} $j$ in ${\game}$,
  then we have $w_j^{(1)}=1=w_j^{(2)}$ due to $\Vert w^{(1)}\Vert_1=\Vert w^{(2)}\Vert_1=1$, so that $w^{(1)}=w^{(2)}$.
  Thus we assume that the number of non-null {\voter}s is at least two and $w_j^{(1)},w_j^{(2)}\in (0,1)$ for all non-null
  {\voter}s $j$ in the following.   
  
  We have $\game(\hat{x})=1$ and $\game(\hat{x}-\varepsilon\cdot e_j)=0$, where 
  $1\le j\le n$ is a non-null {\voter} and $\varepsilon>0$ is arbitrary (but suitably small). Thus we have $q^{(2)}=q^{(1)}$.
  For two arbitrary non-null {\voter}s indices $1\le j_1,j_2\le n$ and suitably small $\varepsilon>0$ we have
  $$
    q^{(1)}=w^{(1)} (\hat{x}-w_{j_2}^{(1)}\cdot \varepsilon\cdot e_{j_1}+w_{j_1}^{(1)}\cdot \varepsilon\cdot e_{j_2})=
    w^{(2)} (\hat{x}-w_{j_2}^{(1)}\cdot \varepsilon\cdot e_{j_1}+w_{j_1}^{(1)}\cdot \varepsilon\cdot e_{j_2})=q^{(2)}.
  $$
  Thus 
  $$
    \frac{w_{j_1}^{(1)}}{w_{j_2}^{(1)}}=\frac{w_{j_1}^{(2)}}{w_{j_2}^{(2)}}
  $$    
  and we conclude $w^{(1)}=w^{(2)}$ from $\Vert w^{(1)}\Vert_1=\Vert w^{(2)}\Vert_1=1$ and $w^{(1)}_i=w^{(2)}_i=$ for
  the null {\voter}s $i$.
\end{Proof}

For the case of quota $q=1$ we remark, that any (normalized) weight vector $w\in[0,1]^n$ leads to the same continuous
threshold game.

Given a finite number $k$ of continuous threshold games $\game_1,\dots,\game_k$ we call the game arising by
$$
  \game(x)=\min_{1\le i\le k} \left\{\game_i(x)\right\}
$$
the \emph{intersection} of the continuous threshold games $\game_i$. $\game$ is indeed a continuous simple game, but not every
continuous simple game can be written as  a finite intersection of continuous threshold games. An example is given by the
continuous simple game $\game$ with $\game(x)=1$ if $\frac{1}{n}\cdot\sum_{i=1}^n x_i^2\ge\frac{1}{2}$ and $\game(x)=0$ 
otherwise. Here infinitely many continuous threshold games are needed in the intersection.  

\begin{Lemma}
  The quota function $q$ of a continuous weighted game $\game:[0,1]^n\rightarrow[0,1]$ is unique. If $q$ is monotone and 
  continuous, then also the weights $w_i$ are unique. 
\end{Lemma}
\begin{Proof}
  Since $\game(x,\dots,x)=q(x)$ for all $x\in[0,1]$ the quota function is uniquely defined.
  
  Since $q(0)=0$, $q(1)=1$, $q$ is monotone and continuous, there exists a value $x\in[0,1]$ such that $y=q^-1(x)\in(0,1)$ is 
  uniquely defined. If $\game(y,\dots,y,y+\varepsilon,y,\dots,y)=\game(y,\dots,y)$ for $\varepsilon=\min(y,1-y)/2$ and
  a modified position $1\le i\le n$, then $w_i=0$. For all other positions we have $w_j>0$. Let $i$ and $j$ be two positions
  with $w_i,w_j>0$. For suitably small $\varepsilon>0$ there exists a unique $\delta>0$ such that
  $\game(z)=game(y,\dots,y)$, where $z_i=y+\varepsilon$, $z_j=y-\delta$, and $z_h=y$ for $h\neq i,j$. From this we
  conclude $\varepsilon w_i=\delta w_j$, so that we can uniquely determine all $w_h$ due to $\Vert w\Vert_1=1$.
\end{Proof}

So, depending on the chosen definition of weightedness, the corresponding weighted representations are either unique or not.
For the different versions of weightedness the connection to the properties proper, strong, and constant-sum is as follows: 

\begin{Lemma}
  All continuous linearly weighted games are proper, strong, and constant-sum.
\end{Lemma}
\begin{Proof}
  For a given continuous weighted game $\game$ let $w\in[0,1]^n$ with $\Vert w\Vert_1=1$ be suitable (normalized) weights.
  With this we have
  $$
    \game(x)+\game(\mathbf{1}-x)=w^Tx+w^T(\mathbf{1}-x)=\Vert w\Vert_1=1
  $$ 
  for all $x\in[0,1]^n$.
\end{Proof}

\begin{Lemma}
  A continuous threshold game $\game$ with (normalized) weights $w$ and quota $q\in(0,1]$ is proper if and only if
  $q>\frac{1}{2}$.
\end{Lemma}
\begin{Proof}
  Since 
  $$
    \game\left(\frac{1}{2}\cdot\mathbf{1}\right)=\frac{1}{2}w^T\mathbf{1}=1-\frac{1}{2}w^T\mathbf{1}
    =\game\left(\mathbf{1}-\frac{1}{2}\cdot\mathbf{1}\right)
  $$
  the game $\game$ is non-proper for $q\le\frac{1}{2}$. Now assume $q>\frac{1}{2}$.
  We have $\game(x)\in\{0,1\}$ for all $x\in[0,1]^n$. Assume that both $\game(x)=1$ and
  $\game(\mathbf{1}-x)=1$. Then we would have $w^Tx\ge q$ and $w^T(\mathbf{1}-x)\ge q$ so that
  $$
    1=\Vert w\Vert_1\ge 2q,
  $$     
  which is a contradiction to $q>\frac{1}{2}$.
\end{Proof}

\begin{Lemma}
  A continuous threshold game $\game$ with (normalized) weights $w$ and quota $q\in(0,1]$ is strong if and only if
  $q\le\frac{1}{2}$.
\end{Lemma}
\begin{Proof}
  Since 
  $$
    \game\left(\frac{1}{2}\cdot\mathbf{1}\right)=\frac{1}{2}w^T\mathbf{1}=1-\frac{1}{2}w^T\mathbf{1}
    =\game\left(\mathbf{1}-\frac{1}{2}\cdot\mathbf{1}\right)
  $$
  the game $\game$ is non-strong for $q>\frac{1}{2}$. Now assume $q\le\frac{1}{2}$.
  We have $\game(x)\in\{0,1\}$ for all $x\in[0,1]^n$. Assume that both $\game(x)=0$ and
  $\game(\mathbf{1}-x)=0$. Then we would have $w^Tx<q$ and $w^T(\mathbf{1}-x)< q$ so that
  $$
    1=\Vert w\Vert_1< 2q,
  $$     
  which is a contradiction to $q\le\frac{1}{2}$.
\end{Proof}

\begin{Corollary}
  No continuous threshold game can be constant-sum.  
\end{Corollary}

\begin{Lemma}
  A continuous weighted game $\game$ with (normalized) weights $w$ and quota function $q:[0,1]\rightarrow[0,1]$
  is proper if and only if $q(y)+q(1-y)\le 1$ for all $y\in[0,1]$.
\end{Lemma}
\begin{Proof}
  For arbitrary $x\in[0,1]^n$ we have
  $$
    \game(x)+\game(\mathbf{1}-x)=q(w^Tx)+q(w^T(\mathbf{1}-x))=q(y)+q(1-y)\le 1,
  $$
  where $y=w^Tx\in[0,1]$.  
\end{Proof}

\begin{Lemma}
  A continuous weighted game $\game$ with (normalized) weights $w$ and quota function $q:[0,1]\rightarrow[0,1]$
  is strong if and only if $q(y)+q(1-y)\ge 1$ for all $y\in[0,1]$.
\end{Lemma}
\begin{Proof}
  For arbitrary $x\in[0,1]^n$ we have
  $$
    \game(x)+\game(\mathbf{1}-x)=q(w^Tx)+q(w^T(\mathbf{1}-x))=q(y)+q(1-y)\ge 1,
  $$
  where $y=w^Tx\in[0,1]$.  
\end{Proof}

\begin{Corollary}
  A continuous weighted game $\game$ with (normalized) weights $w$ and quota function $q:[0,1]\rightarrow[0,1]$
  is constant-sum if and only if $q(y)+q(1-y)=1$ for all $y\in[0,1]$.
\end{Corollary}

The notion of a power index can be transfered as follows:

\begin{Definition}
  \label{def_const_power_index}
  Let $\mathbb{V}_n\subseteq \mathbb{B}_n$ a class of continuous Boolean games consisting of $n$~{\voter}s.
  A power index (on $\mathbb{V}_n$) is a mapping $P:\mathbb{V}_n\rightarrow \mathbb{R}^n$.
\end{Definition}

The four properties of power indices for subclasses of Boolean games, see Definition~\ref{def_nice_power_index_properties},
can be restated one to one for power indices for subclasses of continuous Boolean games. 

Before we give definitions for the Shapley-Shubik index, the absolute Banzhaf index and the nucleolus for continuous simple 
games in Section~\ref{sec_power_indices}, we discuss some special classes of continuous games in the next section. 

\section{Examples of continuous games}
\label{sec_examples_continuous}

The definitions of linearly weighted, threshold, and weighted continuous simple games in the previous section
allow a compact representation of those games given a weight vector $w$ and eventually a quota or quota function $q$.

Numerous theoretical models for the behavior of politicians are based on the so-called median voter model, see e.g.\ 
\cite{downs1957economic,downs1957economicjournal,meltzer1981rational,romer1975individual}. The median voter theorem states that in a voting system
with a single majority decision rule, the most probable elected alternative is the one which is most preferred by the median
voter. The key assumptions of a $1$-dimensional policy space with single peaked preferences are met in our context.
So similarly to Hotelling's law, according to the median voter model, politicians try to adjust their opinions near 
the preferences of the expected median voter. In practice there are several limitations for the median voter theorem
so that the explanatory power of the median voter model is actually rather low, see e.g. \cite{stadelmann2012evaluating}. 

In our context the situation is a bit easier. Given the single peaked preferences $x_i\in[0,1]$ of the voters, 
the aggregated group decision can be any number in $[0,1]$, i.e.\ we neither have to choose within a finite number
of alternatives nor do we indirectly influence future decisions by electing a representative. So it makes quite some
sense to utilize the median aggregation rule given by
\begin{equation}
  \game(x_1,\dots,x_n)=\left\{\begin{array}{rcl}x_{\pi\left(\frac{n+1}{2}\right)}&:&n\equiv 1\pmod 2,\\
  \frac{1}{2}\cdot x_{\pi\left(\frac{n}{2}\right)}+\frac{1}{2}\cdot x_{\pi\left(\frac{n+2}{2}\right)}&:&n\equiv 0\pmod 2,\end{array}\right.
\end{equation} 
where $\pi$ is a permutation such that $x_{\pi(1)}\le\dots\le x_{\pi(n)}$. This decision rule can be slightly generalized by introducing non-negative
weights $w_i$ for all voters $1\le i\le n$ such that $\Vert w\Vert_1=\sum_{i=1}^n w_i>0$. With a permutation $\pi$ as before, 
let $\underline{i}$ be the smallest index such that $\sum_{j=1}^{\underline{i}} w_{\pi(j)}\ge
\Vert w\Vert_1/2$. Similarly, let $\overline{i}$ be the largest index such that $\sum_{j=\overline{i}}^{n} w_{\pi(j)}\ge
\Vert w\Vert_1/2$. If $\underline{i}=\overline{i}$ we set $\game(x_1,\dots,x_n)=x_{\pi\left(\underline{i}\right)}$ and 
$\game(x_1,\dots,x_n)=\left(x_{\pi\left(\underline{i}\right)}+x_{\pi\left(\overline{i}\right)}\right)/2$ otherwise. We call
this procedure the weighted median aggregation rule.

And indeed, continuous games (without our more general notion) equipped with the weighted median aggregation rule are 
e.g.\ studied in \cite{maaser2007equal,maaser2012mean,maaser2012note,napel2004power}. We will see that some
formulas for power indices, defined in the subsequent sections, can be significantly simplified for the (weighted) median
aggregation rule.

Another source of group aggregation rules is the field of opinion dynamics. Assume that each voter starts with an
initial opinion $x_i\in[0,1]$ followed by a dynamic process of exchanging opinions between the individuals. Such
a opinion dynamics influences the initial opinions in a certain way, so that the opinion $x_i'$, after some rounds
of interaction, may significantly differ from the initial ones. Group aggregation for the final opinions $x_i'$
may be performed using weighted voting or the median aggregation rule. Several models for opinion dynamics, i.e.\
specifications how the $x_i$ are modified to the $x_i'$ have been proposed in the literature. Here we only mention
the Lehrer-Wagner model, see e.g. \cite{lehrer1981rational}, the bounded confidence
model, see e.g.\ \cite{hegselmann2002opinion}, model based on opinion leaders, see e.g.\ 
\cite{katz1970personal,van2011measuring}, and the more recent models proposed by Grabisch and Rusinowska 
\cite{grabisch2011model,grabisch2010model,grabisch2011influence,maruani2012and,grabisch2010different} (being based
on the ground of \cite{hoede1982theory}). An overview of social and economic networks is given in \cite{jackson2010social}.

\section{Generalizing three power indices}
\label{sec_power_indices}

In this section we propose generalizations of the Shapley-Shubik index, the Banzhaf index and the nucleolus for 
continuous simple games, which are, in a certain sense, in line with the definitions for simple games or 
$(j,k)$-simple games. We illustrate our definitions by computing the respective indices for the functions
$\hat{\game}(x_1,x_2,x_3)=\frac{1x_1^2+2x_2^2+3x_3^2}{6}$ and $\tilde{\game}(x_1,x_2,x_3)=x_1x_2^2x_3^3$.

\subsection{Shapley-Shubik index}

One interpretation for the definition of the Shapley-Shubik index for simple games is the following:
\begin{enumerate}
  \item[(1)] According to the \textit{veil of ignorance}, the set of vote vectors has no structure, i.e.\ votes 
             are independent and each of the $2^n$ $\{0,1\}$-vectors occurs with equal probability.
  \item[(2)] Assume that the voters are arranged in a sequence and called one by one. After the $i$th voter in the
             current sequence has expressed his vote, an output alternative may be excluded independently from the
             votes of the subsequent voters. Here all sequences are equally probable and the exclusion of an output
             alternative is counted just once, i.e.\ it is counted for the first player who excludes it.
\end{enumerate}             

Going along the same lines for $(j,k)$ simple games, we have $j^n$ possible input vectors in (1) and $n!$ possible
sequences in (2). The notion of an $i$-pivot in Definition~\ref{def_i_pivot} exactly determines the voter who excludes
output alternative $i$ or $i+1$, where we have to consider the direction of the exclusion to avoid double counting.

Lets look at the highest and the lowest possible outcome of a $(j,k)$ simple game $\game$ given the first $i$ votes
$x_1,\dots,x_i$. Due to monotonicity the highest possible outcome occurs if the remaining $n-i$ voters vote for the 
highest possible (input) alternative. Similarly, the lowest possible outcome occurs if the remaining $n-i$ voters vote for the 
lowest possible (input) alternative. For continuous simple games the extremal input alternatives are given by $0$ and $1$ 
so that we define:  

\begin{Definition} $\,$\\[-2mm]
  \begin{itemize}
    \item $\overline{\tau}:[0,1]^n\times \{1,\dots,n\}\rightarrow [0,1]^n$, $(x,i)\mapsto (y_1,\dots,y_n)$,
          where $y_j=x_j$ for all $1\le j\le i$ and $y_j=1$ otherwise;
    \item $\underline{\tau}:[0,1]^n\times \{1,\dots,n\}\rightarrow [0,1]^n$, $(x,i)\mapsto (y_1,\dots,y_n)$,
          where $y_j=x_j$ for all $1\le j\le i$ and $y_j=0$ otherwise.      
  \end{itemize}
\end{Definition}

With this we can \textit{count} the \textit{number of excluded} output alternatives and \textit{sum} over all possible
sequences and vote distributions. Since the output interval $[0,1]$ is continuous, \textit{counting} here means to measure
the length of the newly excluded interval. There are $n!<\infty$ possible sequences of the $n$ voters, so that summing here
really means summing up. Since the space $[0,1]^n$ of possible vote distributions is continuous we have to utilize integrals:

\begin{Definition}
  \label{def_continuous_SSI}
  Let $\game:[0,1]^n\rightarrow[0,1]$ be a continuous simple game. The Shapley-Shubik index $\operatorname{SSI}_i(\game)$ of 
  {\voter}~$i$ in $\game$ is given by
  \begin{eqnarray*}
    &&\frac{1}{n!}\cdot\sum_{\pi\in\mathcal{S}_n} \int_0^1\dots\int_0^1
    \Big(\game\left(\overline{\tau}\!\left(x,\pi^{-1}(i)-1\right)\right)-\game\left(\overline{\tau}\!\left(x,\pi^{-1}(i)\right)\right)\Big)\\
    &&+\Big(\game\left(\underline{\tau}\!\left(x,\pi^{-1}(i)\right)\right)-\game\left(\underline{\tau}\!\left(x,\pi^{-1}(i)-1\right)\right)\Big) 
    \operatorname{d}x_1\,\dots\,\operatorname{d}x_n,
  \end{eqnarray*}
  where $\mathcal{S}_n$ denotes the symmetric group on $n$ elements, i.e.\ the set of permutations or bijections from
  $\{1,\dots,n\}$ to $\{1,\dots,n\}$.
\end{Definition}

For our two examples we obtain
$$
  \operatorname{SSI}(\hat{\game})=\left(\frac{1}{6},\frac{2}{6},\frac{3}{6}\right)
  =\left(0.1\overline{6},0.\overline{3},0.5\right)
$$
and
$$
  \operatorname{SSI}(\tilde{\game})=\left(\frac{35}{144},\frac{50}{144},\frac{59}{144}\right)
  =\left(0.2430\overline{5},0.347\overline{2},0.4097\overline{2}\right).
$$
The detailed computations are stated in Appendix~\ref{appendix_SSI}.

While the \textit{story} of interpreting the Shapley-Shubik index, stated at the beginning of this section, may be considered
to be nice, more serious characterizations involve a so-called axiomatization, see e.g.\ \cite{dubey1975uniqueness}:

\begin{Lemma}
  Let $P:\mathcal{S}_n\rightarrow\mathbb{R}^n$ be a power index. If $P$ satisfies symmetry, efficiency, the null voter
  property, and the transfer axiom, then $P$ coincides with the Shapley-Shubik index. 
\end{Lemma}

In order to define the transfer axiom for simple games we need:

\begin{Definition}
  For two Boolean games $\game_1,\game_2\in\mathcal{S}_n$ we define $\game_1 \vee \game_2$ by 
  $\left(\game_1 \vee \game_2\right)(S)=\max\left(\game_1(S),\game_2(S)\right)$ for all $S\subseteq N$. Similarly, we define 
  $\game_1 \wedge \game_2$ by $\left(\game_1 \wedge \game_2\right)(S)=\min\left(\game_1(S),\game_2(S)\right)$.
\end{Definition}

\begin{Definition}
  \label{def_transfer_simple}
  A power index $P:\mathcal{V}_n\rightarrow\mathbb{R}^n$ satisfies the \textbf{transfer axiom}, if 
  $$
    P(\game_1)+P(\game_2)=P(\game_1\wedge \game_2)+P(\game_1\vee \game_2)
  $$
  for all $\game_1,\game_2\in \mathcal{V}_n$ such that also $\left(\game_1\wedge \game_2\right),\left(\game_1\vee \game_2\right)
  \in\mathcal{V}_n$, where $\mathcal{V}_n$ is a subclass of (binary) Boolean games. 
\end{Definition}

Definition~\ref{def_transfer_simple} can be restated directly for continuous Boolean games using:

\begin{Definition}
  For two continuous Boolean games $\game_1,\game_2$ we define $\game_1 \vee \game_2$ by 
  $\left(\game_1 \vee \game_2\right)(x)=\max\left(\game_1(x),\game_2(x)\right)$ for all $x\in[0,1]^n$. Similarly, we define 
  $\game_1 \wedge \game_2$ by $\left(\game_1 \wedge \game_2\right)(x)=\min\left(\game_1(x),\game_2(x)\right)$.
\end{Definition}

If $\game_1,\game_2\in\mathbb{S}_n$, then also $\left(\game_1\wedge \game_2\right),\left(\game_1\vee \game_2\right)
\in\mathbb{S}_n$. Directly from the definitions we conclude:

\begin{Lemma}  
  The Shapley-Shubik index $\operatorname{SSI}$ is symmetric, positive, and satisfies both the null voter property 
  and the transfer axiom on $\mathbb{S}_n$.
\end{Lemma}

\begin{Conjecture}
  \label{conjecture_SSI_efficient}
  The Shapley-Shubik index for continuous simple games is efficient, i.e.\ $\sum_{i=1}^n \operatorname{SSI}_i(\game)=1$ for
  all $\game\in\mathbb{S}_n$.
\end{Conjecture}

\begin{Conjecture}
  Let $P:\mathbb{S}_n\rightarrow\mathbb{R}^n$ be a power index. If $P$ satisfies symmetry, efficiency, the null voter
  property, and the transfer axiom, then $P$ coincides with the Shapley-Shubik index according to
  Definition~\ref{def_continuous_SSI}. 
\end{Conjecture}

As remarked before, the formula for the Shapley-Shubik index can be simplified for the weighted median aggregation rule.
To this end let $w\in\mathbb{R}^n_{\ge 0}$ be a weight vector with $\Vert w\Vert_1>0$. To avoid technical difficulties 
we assume $\sum_{i\in S} w_i\neq\Vert w\Vert_1/2$ for all $S\subseteq N$, i.e.\ that there is always a unique weighted 
median voter. Without proof we state:

\begin{Lemma}
  \label{lemma_SSI_median}
  The Shapley-Shubik index of the weighted median aggregation rule, according to Definition~\ref{def_continuous_SSI}
  is given by the Shapley-Shubik index of the weighted game $\left[\Vert w\Vert_1/2;w_1,\dots,w_n\right]$. 
\end{Lemma}

We give an example in Appendix~\ref{appendix_SSI_median}.

\subsection{Banzhaf index}

One interpretation for the definition of the Banzhaf index for simple games and $(j,k)$ simple games is the following:
\begin{enumerate}
  \item[(1)] According to the \textit{veil of ignorance}, the set of vote vectors has no structure, i.e.\ votes 
             are independent and each of the $j^n$ $J$-vectors occurs with equal probability.
  \item[(2)] Relevant for the measurement of influence is only the number of $(m,l)$-swings (or swings for simple
             games) for voter~$i$ arising if voter~$i$ shifts his chosen alternative by one.
\end{enumerate}             

For continuous simple games the votes of the voters in $N\backslash\{i\}$ are equally distributed in $[0,1]^{n-1}$, so
that we have to use an $(n-1)$-fold integral. The counting of $(m,l)$-swings for the different possible shifts of
the opinion of voter~$1$ can be condensed to a single expression: Given an ordered $j$-partition $S$, we denote by
$\underline{S}$ the $j$-partition arising from $S$ by setting the vote of voters $i$ to alternative $1$. Similarly, we
denote by $\overline{S}$ the $j$-partition arising from $S$ by setting the vote of voters $i$ to alternative $k$. Then
$\game(\underline{S})-\game(\overline{S})$ counts the number of $(m,l)$-swings for voter $i$ given the preferences of
the other voters in $N\backslash\{i\}$. By dividing by $k-1$ this number is contained in $[0,1]$. For continuous simple
games the lowest possible alternative is $0$ and the highest possible alternative is $1$, so that:

\begin{Definition}
  \label{def_continuous_BZI}
  Let $\game:[0,1]^n\rightarrow[0,1]$ be a continuous simple game. The (absolute) Banzhaf index $\operatorname{BZI}_i(\game)$ of 
  {\voter}~$i$ in $\game$ is given by
  \begin{eqnarray*}
    \!\!\!\!\!&&\int_0^1\dots\int_0^1 \Big(\game(x_1,\dots,x_{i-1},1,x_{i+1},\dots,n)-\game(x_1,\dots,x_{i-1},0,x_{i+1},\dots,n)\Big)\\%\,
    \!\!\!\!\!&&\operatorname{d}x_1\dots\operatorname{d}x_{i-1}\, \operatorname{d}x_{i+1}\dots\operatorname{d}x_n.
  \end{eqnarray*}
\end{Definition}

For the two continuous simple games, introduced at the beginning of this section, we obtain:

\begin{eqnarray*}
  \operatorname{BZI}_1(\hat{\game}) &=& \int_0^1\int_0^1 \left(\frac{1+2x_2^2+3x_3^3}{6}-\frac{0+2x_2^2+3x_3^3}{6}\right)\operatorname{d}x_2\,\operatorname{d}x_3=\frac{1}{6}\\
  \operatorname{BZI}_2(\hat{\game}) &=& \int_0^1\int_0^1 \left(\frac{1x_1+2+3x_3^3}{6}-\frac{1x_1^1+0+3x_3^3}{6}\right)\operatorname{d}x_1\,\operatorname{d}x_3=\frac{2}{6}\\
  \operatorname{BZI}_3(\hat{\game}) &=& \int_0^1\int_0^1 \left(\frac{1x_1^2+2x_2^2+3}{6}-\frac{1x_1^2+2x_2^2+0}{6}\right)\operatorname{d}x_1\,\operatorname{d}x_2=\frac{3}{6}
\end{eqnarray*}

Since $\operatorname{BZI}_1(\hat{\game})+\operatorname{BZI}_2(\hat{\game})+\operatorname{BZI}_3(\hat{\game})=1$ no normalization
is necessary.

\begin{eqnarray*}
  \operatorname{BZI}_1(\tilde{\game}) &=& \int_0^1\int_0^1 \left(x_2^2x_3^3-0\right)\operatorname{d}x_2\,\operatorname{d}x_3=\frac{1}{12}\\
  \operatorname{BZI}_2(\tilde{\game}) &=& \int_0^1\int_0^1 \left(x_1x_3^3-0\right)\operatorname{d}x_1\,\operatorname{d}x_3=\frac{1}{8}\\
  \operatorname{BZI}_3(\tilde{\game}) &=& \int_0^1\int_0^1 \left(x_1x_2^2-0\right)\operatorname{d}x_1\,\operatorname{d}x_2=\frac{1}{6}
\end{eqnarray*}

After normalization we obtain $\frac{1}{9}\cdot\left(2,3,4\right)=\left(0.\overline{2},0.\overline{3},0.\overline{4}\right)$ for the
(relative) Banzhaf index.

An axiomatization of the Banzhaf index for simple games was e.g.\ be given in \cite{dubey1979mathematical}:
 
\begin{Lemma}
  Let $P:\mathcal{S}_n\rightarrow\mathbb{R}^n$ be a power index. If $P$ satisfies symmetry, the null voter
  property, the transfer axiom, and the Banzhaf total power, then $P$ coincides with the Banzhaf index. 
\end{Lemma}

\begin{Definition}
  \label{def_total_power_BZI_simple}
  A power index $P:\mathcal{V}_n\rightarrow\mathbb{R}^n$ satisfies \textbf{Banzhaf total power}, if 
  $$
    \sum_{i=1}^n \operatorname{BZI}_i(\game)=\frac{1}{2^{n-1}}\cdot \sum_{i=1}^n\sum_{S\subseteq N\backslash\{i\}}
    \big(\game(S\cup\{i\})-\game(S)\big)
  $$
  for all $\game\in \mathcal{V}_n$, where $\mathcal{V}_n$ is a subclass of (binary) Boolean games.
\end{Definition}

Definition~\ref{def_total_power_BZI_simple} can be restated directly for continuous Boolean games:

\begin{Definition}
  \label{def_total_power_BZI_cont}
  A power index $P:\mathbb{V}_n\rightarrow\mathbb{R}^n$ satisfies \textbf{Banzhaf total power}, if
  $\Vert \operatorname{BZI}(\game)\Vert_1$ coincides with   
  \begin{eqnarray*}
    \!\!\!\!\!&&\sum_{i=1}^n\int_0^1\dots\int_0^1 \Big(\game(x_1,\dots,x_{i-1},1,x_{i+1},\dots,n)-\game(x_1,\dots,x_{i-1},0,x_{i+1},\dots,n)\Big)\\%\,
    \!\!\!\!\!&&\operatorname{d}x_1\dots\operatorname{d}x_{i-1}\, \operatorname{d}x_{i+1}\dots\operatorname{d}x_n.
  \end{eqnarray*}
  for all $\game\in \mathbb{V}_n$, where $\mathbb{V}_n$ is a subclass of continuous Boolean games.
\end{Definition}

Directly from the definitions we conclude:

\begin{Lemma}  
  The Banzhaf index $\operatorname{BZI}$ is symmetric, positive, and satisfies the null voter property,  
  the transfer axiom, and Banzhaf total power on $\mathbb{S}_n$.
\end{Lemma}

\begin{Conjecture}
  Let $P:\mathbb{S}_n\rightarrow\mathbb{R}^n$ be a power index. If $P$ satisfies symmetry, the null voter
  property, the transfer axiom, and the Banzhaf total power, then $P$ coincides with the Banzhaf index according to
  Definition~\ref{def_continuous_BZI}. 
\end{Conjecture}

\subsection{Nucleolus}

\begin{Definition}
  Given a continuous simple game $\game:[0,1]^n\rightarrow[0,1]$ and a vector $w\in[0,1]^n$ with $\Vert w\Vert_1=1$, 
  the \textbf{excess} of a coalition $x\in[0,1]^n$ is given by $e^\game(x,w)=\game(x)-w^Tx\in[-1,1]$.
\end{Definition}  

The excess vector for the case of simple games is generalized to:

\begin{Definition}
  Given a continuous simple game $\game:[0,1]^n\rightarrow[0,1]$ and a vector $w\in[0,1]^n$ with $\Vert w\Vert_1=1$, the
  \textbf{excess function} is given by $E_{w}^{\game}:[-1,1]\rightarrow[0,1]$, 
  $$
    c\mapsto \operatorname{vol}\left(\left\{x\in[0,1]^n\,:\,\game(x)-w^Tx\ge c\right\}\right),
  $$   
  where $\operatorname{vol}(S)$ denotes the $n$-dimensional volume of a subset $S\subseteq \mathbb{R}^n$. (Here we assume
  that the mapping $\game$ is \textit{regular enough}, e.g.\ piecewise continuous, so that those volumes exist.)
\end{Definition}  

Instead of the lexicographic ordering for two excess vectors we define:

\begin{Definition}
  For two integrable functions $f_1:[-1,1]\rightarrow[0,1]$ and $f_2:[-1,1]\rightarrow[0,1]$, we write $f_1\le f_2$ if
  there exists a constant $c\in[-1,1]$ such that $f_1(y)\le f_2(y)$ for all $y\in [c,1]$ and 
  $\int_{c}^1 f_1(y)\,\operatorname{d}y<\int_{c}^1 f_2(y)\,\operatorname{d}y$.   
\end{Definition} 

\begin{Definition}
  \label{def_nucleolus}
  For a continuous simple game $\game:[0,1]^n\rightarrow[0,1]$ the \textbf{nucleolus} $\mathbf{\operatorname{Nuc}(\game)}$ is given by
  $$
    \left\{w\in[0,1]^n\,:\,\Vert w\Vert_1=1,\, E_{w}^{\game}\le E_{\hat{w}}^{\game}\,\forall \hat{w}\in[0,1]^n: \Vert\hat{w}\Vert_1=1\right\}.
  $$
\end{Definition}

\begin{Conjecture}
  Under mild technical assumptions for a continuous simple game $\game:[0,1]^n\rightarrow[0,1]$, we have $|\operatorname{Nuc}(\game)|\le 1$.
\end{Conjecture} 

By definition the elements of the nucleolus are positive and efficient. Of course we also want to compute the nucleolus for our two
examples. Unfortunately we have no general algorithm at hand, which is capable of solving the optimization problem stated
in Definition~\ref{def_nucleolus}. For $\hat{\game}$ we can compute the nucleolus to be $\frac{1}{6}\cdot(1,2,3)$, i.e.\ it
coincides with the Shapley-Shubik and the Banzhaf index, using a tailored analysis in Appendix~\ref{appendix_nucleolus}.
For $\tilde{\game}$ things seem to be much more complicated without the aid of theoretical results. For the similar
two-voter example $\game(x_1,x_2)=x_1x_2^2$ we compute numeric bounds for the elements in the nucleolus in
Appendix~\ref{appendix_nucleolus}. 
 
\section{Power indices when votes are not equiprobable}
\label{sec_not_equiprobable}

Both the Shapley-Shubik index and the Banzhaf index for simple games, $(j,k)$ simple games, or continuous simple games
are based on the assumption that voters vote independently from each other and that they choose each alternative with equal 
probability. The first assumption is clearly violated in several practical contexts. Here we restrict ourselves to 
situations where this assumption is still met. An equal probability for all possible input alternatives makes a certain sense
for (binary) simple games. Here one can have in mind that the roles of the alternatives are swapped if the proposal is 
logically negated. As argued in e.g.\ \cite{felsenthal1997ternary}, for the special case of $(3,2)$ simple games, where
the central alternative is abstention, things are quite different. In some real-world legislatures, where abstention is 
allowed, the rate of abstention is rather low, while in others it is considerably higher. For the Banzhaf index different
probabilities for the two options were e.g.\ considered in \cite{kaniovski2008exact}.

For continuous simple games we model this more general situation by assuming a density function $f_i$, i.e.\ 
$f_i:[0,1]\rightarrow\mathbb{R}_{\ge 0}$ with $\int_0^1 f_i(x)\,\operatorname{d}x=1$, for each voter $i\in N$. 
With this we propose: 

\begin{Definition}
  \label{def_continuous_SSI_non_equiprobable}
  Let $\game:[0,1]^n\rightarrow[0,1]$ be a continuous simple game. The density Shapley-Shubik index $\operatorname{SSI}^f_i(\game)$ of 
  {\voter}~$i$ in $\game$ is given by
  \begin{eqnarray*}
    &&\frac{1}{n!}\cdot\sum_{\pi\in\mathcal{S}_n} \int_0^1\dots\int_0^1
    \Big(\Big(\game\left(\overline{\tau}\!\left(x,\pi^{-1}(i)-1\right)\right)-\game\left(\overline{\tau}\!\left(x,\pi^{-1}(i)\right)\right)\Big)\\
    &&+\Big(\game\left(\underline{\tau}\!\left(x,\pi^{-1}(i)\right)\right)-\game\left(\underline{\tau}\!\left(x,\pi^{-1}(i)-1\right)\right)\Big)\Big) 
    \cdot f_1(x_1)\dots f_n(x_n)\,\operatorname{d}x_1\,\dots\,\operatorname{d}x_n,
  \end{eqnarray*}
  where $\mathcal{S}_n$ denotes the symmetric group on $n$ elements, i.e.\ the set of permutations or bijections from
  $\{1,\dots,n\}$ to $\{1,\dots,n\}$ and $f=(f_1,\dots,f_n)$ is a vector of density functions.
\end{Definition}
 
For the special case of the weighted median aggregation rule this definition was (in its simplified version) e.g.\ used in 
\cite{kurz2012egalitarian}. As a small example we consider the median aggregation rule for a continuous simple game $\game$ with
three voters and density functions, which are given by 
\begin{eqnarray*}
  f_1(x)&=&\frac{3}{4}\cdot\left(1-x^2\right),\\
  f_2(x)&=&\frac{3}{8}\cdot\left(1+x^2\right), \text{and}\\
  f_3(x)&=&\frac{3}{8}\cdot\left(1+x^2\right),
\end{eqnarray*}  
for $x\in[-1,1]$ and zero otherwise. We can easily check that the three stated functions are indeed density functions. With this we have
\begin{eqnarray*}
  \operatorname{SSI}^f_1(\game)&=&\int_{-1}^1 \int_{-1}^{x_1}\int_{x_1}^1 f(x)\,\operatorname{d}x_2,\operatorname{d}x_3\, \operatorname{d}x_1+\int_{-1}^1 \int_{-1}^{x_1}\int_{x_1}^1 f(x)\,\operatorname{d}x_3,\operatorname{d}x_2\, \operatorname{d}x_1\\
               &=& \frac{554}{13440}\approx 0.04122 \\
  \operatorname{SSI}^f_2(\game)&=&\int_{-1}^1 \int_{-1}^{x_2}\int_{x_2}^1 f(x)\,\operatorname{d}x_1,\operatorname{d}x_3\, \operatorname{d}x_2+\int_{-1}^1 \int_{-1}^{x_2}\int_{x_2}^1 f(x)\,\operatorname{d}x_3,\operatorname{d}x_1\, \operatorname{d}x_2\\
               &=& \frac{563}{13440}\approx 0.04189 \\
  \operatorname{SSI}^f_3(\game)&=&\int_{-1}^1 \int_{-1}^{x_3}\int_{x_3}^1 f(x)\,\operatorname{d}x_1,\operatorname{d}x_2\, \operatorname{d}x_3+\int_{-1}^1 \int_{-1}^{x_3}\int_{x_3}^1 f(x)\,\operatorname{d}x_2,\operatorname{d}x_1\, \operatorname{d}x_3\\
               &=& \frac{563}{13440}\approx 0.04189 \\                         
\end{eqnarray*} 
where we use the abbreviation $f(x)=f_1(x_1)\cdot f_2(x_2)\cdot f_3(x_3)$.

\begin{Definition}
  \label{def_continuous_BZI_non_equiprobable}
  Let $\game:[0,1]^n\rightarrow[0,1]$ be a continuous simple game. The density (absolute) Banzhaf index $\operatorname{BZI}^f_i(\game)$ of 
  {\voter}~$i$ in $\game$ is given by
  \begin{eqnarray*}
    \!\!\!\!\!&&\int_0^1\dots\int_0^1 \Big(\game(x_1,\dots,x_{i-1},1,x_{i+1},\dots,n)-\game(x_1,\dots,x_{i-1},0,x_{i+1},\dots,n)\Big)\\%\,
    \!\!\!\!\!&&\cdot f_1(x_1)\dots f_{i-1}(x_{i-1})\cdot f_{i+1}(x_{i+1})\dots f_n(x_n)\,\operatorname{d}x_1\dots\operatorname{d}x_{i-1}\, \operatorname{d}x_{i+1}\dots\operatorname{d}x_n,
  \end{eqnarray*}
  where $f=(f_1,\dots,f_n)$ is a vector of density functions.
\end{Definition}
 
\section{Conclusion}
\label{sec_conclusion}

Measurement of voting power is relevant in many practical applications. The widely used binary voting model does not fit
for several economic problems like e.g.\ tax rates or spending. Here we have proposed some definitions for continuous games and
highlighted their similarity to the corresponding definitions for simple or $(j,k)$ simple games. Some first few assertions, 
known to be true for simple games, are proven to be valid for our new generalized definitions. We do not claim that we have 
found the ultimate truth, but want to stimulate the research for the \textit{right} generalization by presenting our 
educated guess. It is a major task for the future, to transfer known results for simple games for continuous simple games 
and eventually modify our definitions if they do not seem to fit well for a majority of those results. The possibly 
\textit{weakest} part of our suggestions are the generalizations of weightedness. But here the situation even has not been
resolved convincingly for $(j,k)$ simple games. A good benchmark for the proposed definitions of certain properties for
simple, $(j,k)$ simple, and continuous simple games would be, if the version for simple games arises as a specialization to
$(2,2)$ simple games, and the version for continuous simple games arises by taking the limit $j,k\to\infty$.

For three power indices from cooperative game theory we have proposed generalizations for continuous simple games and started
to study their properties. A litmus test might be to check whether those defined power indices can be axiomatized in a similar
fashion than their binary counterparts. For the Shapley-Shubik and the Banzhaf index we have conjectured such
axiomatizations. The key to a possible proof of those conjectures might be a generalized definition of unanimity games.   

We have illustrated our generalized power indices by computing the respective values for several examples. For some parameterized 
classes of such examples it is indeed possible to write down easy formulas, which will be delayed to a more technical follow-up paper. 
For the proposed generalization of the nucleolus even an algorithmic way to compute the corresponding set is missing. Maybe
it makes also sense to consider generalizations for other of the known power indices for simple games. 

We really hope that this paper can partially contribute to the development of a unified framework for measuring decision power 
and stimulates further research in that direction.

%\bibliographystyle{amsplain}
%\bibliography{continous_model}

\providecommand{\bysame}{\leavevmode\hbox to3em{\hrulefill}\thinspace}
\providecommand{\MR}{\relax\ifhmode\unskip\space\fi MR }
\providecommand{\MRhref}[2]{%
  \href{http://www.ams.org/mathscinet-getitem?mr=#1}{#2}
}
\providecommand{\href}[2]{#2}

\appendix

\section{Determining the $\mathbf{\operatorname{SSI}}$ for two continouus simple games}
\label{appendix_SSI}

For the two continuous simple games $\hat{\game}$ and $\tilde{\game}$ from Section~\ref{sec_power_indices} we 
compute the Shapley-Shubik indices. In tables~\ref{table_ex_1_ssi_1}-\ref{table_ex_1_ssi_3}
we give the respective summands for each permutation $\pi\in\mathcal{S}_3$ for $\hat{\game}$. Summarizing the results
we obtain $$\operatorname{SSI}(\hat{\game})=\left(\frac{1}{6},\frac{2}{6},\frac{3}{6}\right).$$

\begin{table}[htp]
  \begin{center}
    \begin{tabular}{cc}
      \hline
      $\pi\in\mathcal{S}_3$ & $3$-fold integral\\
      \hline
      $(1,2,3)$ & $\int\limits_{x\in[0,1]^3} \left(\frac{6}{6}-\frac{x_1^2+5}{6}+\frac{x_1^2}{6}-\frac{0}{6}\right)\,\operatorname{d}x=\frac{1}{6}$ \\
      $(1,3,2)$ & $\int\limits_{x\in[0,1]^3} \left(\frac{6}{6}-\frac{x_1^2+5}{6}+\frac{x_1^2}{6}-\frac{0}{6}\right)\,\operatorname{d}x=\frac{1}{6}$ \\
      $(2,1,3)$ & $\int\limits_{x\in[0,1]^3} \left(\frac{2x_2^2+4}{6}-\frac{x_1^2+2x_2^2+3}{6}+\frac{x_1^2+2x_2^2}{6}-\frac{2x_2^2}{6}\right)\,\operatorname{d}x=\frac{1}{6}$ \\
      $(2,3,1)$ & $\int\limits_{x\in[0,1]^3} \left(\frac{2x_2^2+3x_3^2+1}{6}-\frac{x_1^2+2x_2^2+3x_3^2}{6}+\frac{x_1^2+2x_2^2+3x_3^2}{6}-\frac{2x_2^2+3x_3^2}{6}\right)\,\operatorname{d}x=\frac{1}{6}$ \\
      $(3,1,2)$ & $\int\limits_{x\in[0,1]^3} \left(\frac{3x_3^2+3}{6}-\frac{x_1^2+3x_3^2+2}{6}+\frac{x_1^2+3x_3^2}{6}-\frac{3x_3^2}{6}\right)\,\operatorname{d}x=\frac{1}{6}$ \\
      $(3,2,1)$ & $\int\limits_{x\in[0,1]^3} \left(\frac{2x_2^2+3x_3^2+1}{6}-\frac{x_1^2+2x_2^2+3x_3^2}{6}+\frac{x_1^2+2x_2^2+3x_3^2}{6}-\frac{2x_2^2+3x_3^2}{6}\right)\,\operatorname{d}x=\frac{1}{6}$ \\
      \hline
    \end{tabular}
    \caption{$\operatorname{SSI}_1(\hat{\game})$ for $\hat{\game}(x_1,x_2,x_3)=\frac{1}{6}\cdot(1x_1^2+2x_2^2+3x_3^2)$.}
    \label{table_ex_1_ssi_1}
  \end{center}
\end{table}

\begin{table}[htp]
  \begin{center}
    \begin{tabular}{cc}
      \hline
      $\pi\in\mathcal{S}_3$ & $3$-fold integral\\
      \hline
      $(2,1,3)$ & $\int\limits_{x\in[0,1]^3} \left(\frac{6}{6}-\frac{x_2^2+4}{6}+\frac{x_2^2}{6}-\frac{0}{6}\right)\,\operatorname{d}x=\frac{2}{6}$ \\
      $(2,3,1)$ & $\int\limits_{x\in[0,1]^3} \left(\frac{6}{6}-\frac{x_2^2+4}{6}+\frac{x_2^2}{6}-\frac{0}{6}\right)\,\operatorname{d}x=\frac{2}{6}$ \\
      $(1,2,3)$ & $\int\limits_{x\in[0,1]^3} \left(\frac{x_1^2+5}{6}-\frac{x_1^2+2x_2^2+3}{6}+\frac{x_1^2+2x_2^2}{6}-\frac{x_1^2}{6}\right)\,\operatorname{d}x=\frac{2}{6}$ \\
      $(3,2,1)$ & $\int\limits_{x\in[0,1]^3} \left(\frac{3x_3^2+3}{6}-\frac{2x_2^2+3x_3^2+1}{6}+\frac{2x_2^2+3x_3^2}{6}-\frac{3x_3^2}{6}\right)\,\operatorname{d}x=\frac{2}{6}$ \\
      $(1,3,2)$ & $\int\limits_{x\in[0,1]^3} \left(\frac{x_1^2+3x_3^2+2}{6}-\frac{x_1^2+2x_2^2+3x_3^2}{6}+\frac{x_1^2+2x_2^2+3x_3^2}{6}-\frac{x_1^2+3x_3^2}{6}\right)\,\operatorname{d}x=\frac{2}{6}$ \\
      $(3,1,2)$ & $\int\limits_{x\in[0,1]^3} \left(\frac{x_1^2+3x_3^2+2}{6}-\frac{x_1^2+2x_2^2+3x_3^2}{6}+\frac{x_1^2+2x_2^2+3x_3^2}{6}-\frac{x_1^2+3x_3^2}{6}\right)\,\operatorname{d}x=\frac{2}{6}$ \\
      \hline
    \end{tabular}
    \caption{$\operatorname{SSI}_2(\hat{\game})$ for $\hat{\game}(x_1,x_2,x_3)=\frac{1}{6}\cdot(1x_1^2+2x_2^2+3x_3^2)$.}
    \label{table_ex_1_ssi_2}
  \end{center}
\end{table}

\begin{table}[htp]
  \begin{center}
    \begin{tabular}{cc}
      \hline
      $\pi\in\mathcal{S}_3$ & $3$-fold integral\\
      \hline
      $(3,1,2)$ & $\int\limits_{x\in[0,1]^3} \left(\frac{6}{6}-\frac{3x_3^2+3}{6}+\frac{3x_3^2}{6}-\frac{0}{6}\right)\,\operatorname{d}x=\frac{3}{6}$ \\
      $(3,2,1)$ & $\int\limits_{x\in[0,1]^3} \left(\frac{6}{6}-\frac{3x_3^2+3}{6}+\frac{3x_3^2}{6}-\frac{0}{6}\right)\,\operatorname{d}x=\frac{3}{6}$ \\
      $(1,3,2)$ & $\int\limits_{x\in[0,1]^3} \left(\frac{x_1^2+5}{6}-\frac{x_1^2+3x_3^2+2}{6}+\frac{x_1^2+3x_3^2}{6}-\frac{x_1^2}{6}\right)\,\operatorname{d}x=\frac{3}{6}$ \\
      $(2,3,1)$ & $\int\limits_{x\in[0,1]^3} \left(\frac{2x_2^2+4}{6}-\frac{2x_2^2+3x_3^2+1}{6}+\frac{2x_2^2+3x_3^2}{6}-\frac{2x_2^2}{6}\right)\,\operatorname{d}x=\frac{3}{6}$ \\
      $(1,2,3)$ & $\int\limits_{x\in[0,1]^3} \left(\frac{x_1^2+2x_2^2+3}{6}-\frac{x_1^2+2x_2^2+3x_3^2}{6}+\frac{x_1^2+2x_2^2+3x_3^2}{6}-\frac{x_1^2+2x_2^2}{6}\right)\,\operatorname{d}x=\frac{3}{6}$ \\
      $(2,1,3)$ & $\int\limits_{x\in[0,1]^3} \left(\frac{x_1^2+2x_2^2+3}{6}-\frac{x_1^2+2x_2^2+3x_3^2}{6}+\frac{x_1^2+2x_2^2+3x_3^2}{6}-\frac{x_1^2+2x_2^2}{6}\right)\,\operatorname{d}x=\frac{3}{6}$ \\
      \hline
    \end{tabular}
    \caption{$\operatorname{SSI}_3(\hat{\game})$ for $\hat{\game}(x_1,x_2,x_3)=\frac{1}{6}\cdot(1x_1^2+2x_2^2+3x_3^2)$.}
    \label{table_ex_1_ssi_3}
  \end{center}
\end{table}

In tables~\ref{table_ex_2_ssi_1}-\ref{table_ex_2_ssi_3} we give the respective summands for each permutation
$\pi\in\mathcal{S}_3$ for $\tilde{\game}$. Summarizing the results we obtain 
$$\operatorname{SSI}(\tilde{\game})=\left(\frac{35}{144},\frac{50}{144},\frac{59}{144}\right)
=\left(0.2430\overline{5},0.347\overline{2},0.4097\overline{2}\right).$$

\begin{table}[htp]
  \begin{center}
    \begin{tabular}{cc}
      \hline
      $\pi\in\mathcal{S}_3$ & $3$-fold integral\\
      \hline
      $(1,2,3)$ & $\int\limits_{x\in[0,1]^3} \left(1-x_1+0-0\right)\,\operatorname{d}x=\frac{1}{2}$ \\
      $(1,3,2)$ & $\int\limits_{x\in[0,1]^3} \left(1-x_1+0-0\right)\,\operatorname{d}x=\frac{1}{2}$ \\
      $(2,1,3)$ & $\int\limits_{x\in[0,1]^3} \left(x_2^2-x_1x_2^2+0-0\right)\,\operatorname{d}x=\frac{1}{6}$ \\
      $(2,3,1)$ & $\int\limits_{x\in[0,1]^3} \left(x_3^3-x_1x_3^3+0-0\right)\,\operatorname{d}x=\frac{1}{8}$ \\
      $(3,1,2)$ & $\int\limits_{x\in[0,1]^3} \left(x_2^2x_3^3-x_1x_2^2x_3^3+x_1x_2^2x_3^3-0\right)\,\operatorname{d}x=\frac{1}{12}$ \\
      $(3,2,1)$ & $\int\limits_{x\in[0,1]^3} \left(x_2^2x_3^3-x_1x_2^2x_3^3+x_1x_2^2x_3^3-0\right)\,\operatorname{d}x=\frac{1}{12}$ \\
      \hline
    \end{tabular}
    \caption{$\operatorname{SSI}_1(\tilde{\game})$ for $\tilde{\game}(x_1,x_2,x_3)=x_1x_2^2x_3^3$.}
    \label{table_ex_2_ssi_1}
  \end{center}
\end{table}

\begin{table}[htp]
  \begin{center}
    \begin{tabular}{cc}
      \hline
      $\pi\in\mathcal{S}_3$ & $3$-fold integral\\
      \hline
      $(1,2,3)$ & $\int\limits_{x\in[0,1]^3} \left(x_1-x_1x_2^2+0-0\right)\,\operatorname{d}x=\frac{1}{3}$ \\
      $(1,3,2)$ & $\int\limits_{x\in[0,1]^3} \left(x_1x_3^3-x_1x_2^2x_3^3+x_1x_2^2x_3^3-0\right)\,\operatorname{d}x=\frac{1}{8}$ \\
      $(2,1,3)$ & $\int\limits_{x\in[0,1]^3} \left(1-x_2^2+0-0\right)\,\operatorname{d}x=\frac{2}{3}$ \\
      $(2,3,1)$ & $\int\limits_{x\in[0,1]^3} \left(1-x_2^2+0-0\right)\,\operatorname{d}x=\frac{2}{3}$ \\
      $(3,1,2)$ & $\int\limits_{x\in[0,1]^3} \left(x_1x_3^3-x_1x_2^2x_3^3+x_1x_2^2x_3^3-0\right)\,\operatorname{d}x=\frac{1}{8}$ \\
      $(3,2,1)$ & $\int\limits_{x\in[0,1]^3} \left(x_3^3-x_2^2x_3^3+0-0\right)\,\operatorname{d}x=\frac{1}{6}$ \\
      \hline
    \end{tabular}
    \caption{$\operatorname{SSI}_2(\tilde{\game})$ for $\tilde{\game}(x_1,x_2,x_3)=x_1x_2^2x_3^3$.}
    \label{table_ex_2_ssi_2}
  \end{center}
\end{table}

\begin{table}[htp]
  \begin{center}
    \begin{tabular}{cc}
      \hline
      $\pi\in\mathcal{S}_3$ & $3$-fold integral\\
      \hline
      $(1,2,3)$ & $\int\limits_{x\in[0,1]^3} \left(x_1x_2^2-x_1x_2^2x_3^3+x_1x_2^2x_3^3-0\right)\,\operatorname{d}x=\frac{1}{6}$ \\
      $(1,3,2)$ & $\int\limits_{x\in[0,1]^3} \left(x_1-x_1x_3^3+0-0\right)\,\operatorname{d}x=\frac{3}{8}$ \\
      $(2,1,3)$ & $\int\limits_{x\in[0,1]^3} \left(x_1x_2^2-x_1x_2^2x_3^3+x_1x_2^2x_3^3-0\right)\,\operatorname{d}x=\frac{1}{6}$ \\
      $(2,3,1)$ & $\int\limits_{x\in[0,1]^3} \left(x_2^2-x_2^2x_3^3+0-0\right)\,\operatorname{d}x=\frac{1}{4}$ \\
      $(3,1,2)$ & $\int\limits_{x\in[0,1]^3} \left(1-x_3^3+0-0\right)\,\operatorname{d}x=\frac{3}{4}$ \\
      $(3,2,1)$ & $\int\limits_{x\in[0,1]^3} \left(1-x_3^3+0-0\right)\,\operatorname{d}x=\frac{3}{4}$ \\      
      \hline
    \end{tabular}
    \caption{$\operatorname{SSI}_3(\tilde{\game})$ for $\tilde{\game}(x_1,x_2,x_3)=x_1x_2^2x_3^3$.}
    \label{table_ex_2_ssi_3}
  \end{center}
\end{table}

\section{Determining the $\operatorname{SSI}$ for a weighted median aggregation rule}
\label{appendix_SSI_median}

In order to illustrate Lemma~\ref{lemma_SSI_median} we compute the Shapley-Shubik index of a continuous simple voting game,
which is given as the weighted median. We consider a case of $4$ {\voter}s with weights $5$, $3$, $2$, $1$. The situation
may be described as weighted binary game $[6;5,3,2,1]=[3;2,1,1,1]$, so that we may also consider the weights $2$, $1$, $1$,
and $1$. Since no subset of the weights sums to exactly half of the weight sum the median is uniquely determined in every case.
One can easily compute that $\operatorname{SSI}([3;2,1,1,1])=\left(\frac{1}{2},\frac{1}{6},\frac{1}{6},\frac{1}{6}\right)$.
In the subsequent subsections we will compute the Shapley-Shubik index directly using Definition~\ref{def_continuous_SSI}, which
is indeed a somewhat lengthy computation.

\subsection{Shapley-Shubik power for {\voter} $2$}
\label{subsec_SSI_median_2}

\subsubsection{$\pi\in\Big\{(2,1,3,4),(2,1,4,3),(2,3,1,4),(2,3,4,1),(2,4,1,3),(2,4,3,1)\Big\}$:} For permutations $\pi$, where $\pi(1)=2$,
we have $\game\left(\overline{\tau}(x,\pi^{-1}(2)-1)\right)=1$, $\game\left(\overline{\tau}(x,\pi^{-1}(2))\right)=1$,
$\game\left(\underline{\tau}(x,\pi^{-1}(2))\right)=0$, and $\game\left(\underline{\tau}(x,\pi^{-1}(2)-1)\right)=0$, so that   
the value of the $4$-fold integral is given by $0$.

\subsubsection{$\pi\in\Big\{(3,2,1,4),(3,2,4,1),(4,2,1,3),(4,2,3,1)\Big\}$:} For permutations $\pi$, where $\pi(2)=2$ and
$\pi(1)\neq 1$, we have $\game\left(\overline{\tau}(x,\pi^{-1}(2)-1)\right)=1$, $\game\left(\overline{\tau}(x,\pi^{-1}(2))\right)=1$,
$\game\left(\underline{\tau}(x,\pi^{-1}(2))\right)=0$, and $\game\left(\underline{\tau}(x,\pi^{-1}(2)-1)\right)=0$, so that   
the value of the $4$-fold integral is given by $0$.

\subsubsection{$\pi=(1,2,3,4)$, $\pi=(1,2,4,3)$:}
For the permutations $\pi=(1,2,3,4)$ and $\pi=(1,2,4,3)$ we have $\game\left(\overline{\tau}(x,\pi^{-1}(2)-1)\right)=1$ and 
$\game\left(\underline{\tau}(x,\pi^{-1}(2)-1)\right)=0$. If $x_1\le x_2$ then $\game\left(\overline{\tau}(x,\pi^{-1}(2))\right)=x_2$
and $\game\left(\underline{\tau}(x,\pi^{-1}(2))\right)=x_1$. Similarly, if I $x_1\ge x_2$ then 
$\game\left(\overline{\tau}(x,\pi^{-1}(2))\right)=x_1$ and $\game\left(\underline{\tau}(x,\pi^{-1}(2))\right)=x_2$. Thus
for these two permutations the value of the $4$-fold integral is given by
\begin{eqnarray}
  &&\int_0^1\left(1-\left(\int_0^{x_2}x_2\,\operatorname{d}x_1+  \int_{x_2}^1 x_1\,\operatorname{d}x_1\right)+
  \left(\int_0^{x_2}x_1\,\operatorname{d}x_1+  \int_{x_2}^1 x_2\,\operatorname{d}x_1\right)-0\right)\operatorname{d}x_2\nonumber\\
  &=& \int_0^1 \frac{1}{2}+x_2(1-x_2)\,\operatorname{d}x_2=\frac{2}{3}
\end{eqnarray}     

\subsubsection{$\pi\in\Big\{(1,3,2,4),(1,4,2,3),(3,1,2,4),(4,1,2,3)\Big\}$:} For the permutation $\pi=(1,3,2,4)$ we consider the six different strict orderings of
$x_1$, $x_2$, and $x_3$ in Table~\ref{table_median_SSI_2_1}. The respective integrals over $x_1$, $x_3$, and $x_4$ are
stated in Table~\ref{table_median_SSI_2_2}.

\begin{table}[htp]
  \begin{center}
    \begin{tabular}{ccccc}
      \hline
      ordering & $\!\!\!\!\!\game\left(\overline{\tau}(x,\pi^{-1}(2)-1)\right)\!\!\!$ & $\!\!\!\game\left(\overline{\tau}(x,\pi^{-1}(2))\right)\!\!\!$ &
      $\!\!\game\left(\underline{\tau}(x,\pi^{-1}(2))\right)\!\!$ & $\!\!\game\left(\underline{\tau}(x,\pi^{-1}(2)-1)\right)\!\!$\\
      \hline
      $x_1\!\!<\!\!x_2\!\!<\!\!x_3$ & $x_3$ & $x_2$ & $x_1$ & $x_1$ \\
      $x_1\!\!<\!\!x_3\!\!<\!\!x_2$ & $x_3$ & $x_3$ & $x_1$ & $x_1$ \\
      $x_2\!\!<\!\!x_1\!\!<\!\!x_3$ & $x_3$ & $x_1$ & $x_1$ & $x_1$ \\
      $x_2\!\!<\!\!x_3\!\!<\!\!x_1$ & $x_1$ & $x_1$ & $x_3$ & $x_3$ \\
      $x_3\!\!<\!\!x_1\!\!<\!\!x_2$ & $x_1$ & $x_1$ & $x_1$ & $x_3$ \\
      $x_3\!\!<\!\!x_2\!\!<\!\!x_1$ & $x_1$ & $x_1$ & $x_2$ & $x_3$ \\
      \hline
    \end{tabular}
    \caption{Values of $\game$ for $\pi=(1,3,2,4)$.}
    \label{table_median_SSI_2_1}
  \end{center}
\end{table} 

\begin{table}[htp]
  \begin{center}
    \begin{tabular}{ll}
      \hline
      ordering & $\int_{(x_1,x_3,x_4)}\star$ \\
      \hline
      $x_1<x_2<x_3$ & $\int_{0}^{x_2}\int_{x_2}^1\left(x_3-x_2+x_1-x_1\right) \operatorname{d}x_3\,\operatorname{d}x_1=\frac{x_2^3-2x_2^2+x_2}{2}$ \\
      $x_1<x_3<x_2$ & $\int_{0}^{x_2}\int_{x_1}^{x_2}\left(x_3-x_3+x_1-x_1\right)\operatorname{d}x_3\,\operatorname{d}x_1=0$ \\
      $x_2<x_1<x_3$ & $\int_{x_2}^{1}\int_{x_1}^{1}\left(x_3-x_1+x_1-x_1\right)\operatorname{d}x_3\,\operatorname{d}x_1=\frac{-x_2^3+3x_2^2-3x_2+1}{6}$ \\
      $x_2<x_3<x_1$ & $\int_{x_2}^{1}\int_{x_3}^{1}\left(x_1-x_1+x_3-x_3\right)\operatorname{d}x_1\,\operatorname{d}x_3=0$ \\
      $x_3<x_1<x_2$ & $\int_{0}^{x_2}\int_{x_3}^{x_2}\left(x_1-x_1+x_1-x_3\right)\operatorname{d}x_1\,\operatorname{d}x_3=\frac{x_2^3}{6}$ \\
      $x_3<x_2<x_1$ & $\int_{0}^{x_2}\int_{x_2}^{1}\left(x_1-x_1+x_2-x_3\right)\operatorname{d}x_1\,\operatorname{d}x_3=\frac{x_2^2-x_2^3}{2}$ \\
      \hline
    \end{tabular}
    \caption{Auxiliary integrals for $\pi=(1,3,2,4)$.}
    \label{table_median_SSI_2_2}
  \end{center}
\end{table}

Summing the right hand sides of the rows of Table~\ref{table_median_SSI_2_2} and integrating over $x_2\in[0,1]$ yields
\begin{equation}
  \int_{0}^1 \frac{1}{6}\,\operatorname{d}x_2=\frac{1}{6}.
\end{equation}
Due to symmetry we obtain the same value for $\pi=(1,4,2,3)$, $\pi=(3,1,2,4)$, and $\pi=(4,1,2,3)$.

\subsubsection{$\pi=(3,4,2,1)$, $\pi=(4,3,2,1)$:} For the permutations $\pi=(3,4,2,1)$ and $\pi=(4,3,2,1)$ we have 
$\game\left(\overline{\tau}(x,\pi^{-1}(2)-1)\right)=1$, $\game\left(\overline{\tau}(x,\pi^{-1}(2)-1)\right)=\max(x_2,x_3,x_4)$,
$\game\left(\underline{\tau}(x,\pi^{-1}(2)-1)\right)=0$, and $\game\left(\underline{\tau}(x,\pi^{-1}(2))\right)=\min(x_2,x_3,x_4)$.
Thus
for these two permutations the value of the $4$-fold integral is given by
\begin{equation}
  \int_0^1\int_0^1\int_0^1\left(1-\max(x_2,x_3,x_4)+\min(x_2,x_3,x_4)\right)\operatorname{d}x_2\,\operatorname{d}x_3\,\operatorname{d}x_4=\frac{1}{2}.
\end{equation}     

\subsubsection{$\pi\in\Big\{(1,3,4,2),(1,4,3,2),(3,1,4,2),(4,1,3,2),(3,4,1,2),(4,3,1,2)\Big\}$:} For permutations $\pi$ with
$\pi(4)=2$, we consider the three cases whether the maximum and minimum of $x_1,x_3,x_4$ is equal to $x_1$ or another number.
The respective values of $\game$ are stated in Table~\ref{table_median_SSI_2_3}, where we use the abbreviation $x'=(x_1,x_3,x_4)$.

\begin{table}[htp]
  \begin{center}
    \begin{tabular}{cccc}
      \hline
      $\max x'$ & $\min x'$ & $\game\left(\overline{\tau}(x,\pi^{-1}(2)-1)\right)$ & $\game\left(\overline{\tau}(x,\pi^{-1}(2))\right)$\\
      \hline
      $x_1$ & $\neq x_1$      & $x_1$ & $\!\!\!\!\!\!\!\!\!\!\left\{\!\!\!\begin{array}{rcl}\max(x_2,x_3,x_4)&\!\!\!:\!\!\!&x_2\le x_1\\x_1&\!\!\!:\!\!\!&x_2\ge x_1\end{array}\right.$ \\
      $\neq x_1$ & $x_1$      & $\min(x_3,x_4)$ & $\min(x_2,x_3,x_4)$ \\
      $\neq x_1$ & $\neq x_1$ & $x_1$ & $x_1$ \\
      \hline
      $\max x'$ & $\min x'$ & $\game\left(\underline{\tau}(x,\pi^{-1}(2))\right)$ & $\game\left(\underline{\tau}(x,\pi^{-1}(2)-1)\right)$\\
      \hline
      $x_1$ & $\neq x_1$      & $\max(x_2,x_3,x_4)$ & $\max(x_3,x_4)$ \\
      $\neq x_1$ & $x_1$      & $\!\left\{\!\!\!\begin{array}{rcl}\min(x_2,x_3,x_4)&\!\!\!:\!\!\!&x_2\ge x_1\\x_1&\!\!\!:\!\!\!&x_2\le x_1\end{array}\right.$ & $x_1$\\
      $\neq x_1$ & $\neq x_1$ & $x_1$ & $x_1$ \\
      \hline
    \end{tabular}
    \caption{Values of $\game$ for permutations $\pi$ with $\pi(4)=2$.}
    \label{table_median_SSI_2_3}
  \end{center}
\end{table}

By looking at the $24$ strict orderings of $x_1,x_2,x_3,x_4$ we compute the $4$-fold to be $\frac{1}{6}$. (For $4$ orderings
we obtain a value of $\frac{1}{60}$, for $12$ orderings a value of $\frac{1}{120}$, and for $8$ orderings a value of $0$.)

\subsubsection{Summarizing the $24$ permutations for $\operatorname{SSI}_2(\game)$}
$$
  \operatorname{SSI}_2(\game)=\frac{1}{4!}\cdot\left(6\cdot 0+4\cdot 0+2\cdot\frac{2}{3}+4\cdot\frac{1}{6}+2\cdot\frac{1}{2}
  +6\cdot\frac{1}{6}\right)=\frac{4}{24}=\frac{1}{6}.
$$

\subsection{Shapley-Shubik power for {\voter} $3$ and {\voter} $4$}

Since {\voter}s $3$ and $4$ behave as {voter}~$2$, i.e.\ they are symmetric, in the weighted median, we obtain 
$\operatorname{SSI}_3(\game)=\operatorname{SSI}_4(\game)=\frac{1}{6}$.

\subsection{Shapley-Shubik power for {\voter} $1$}

Assuming Conjecture~\ref{conjecture_SSI_efficient} we would obtain
$$
  \operatorname{SSI}_1(\game)=1-\operatorname{SSI}_2(\game)-\operatorname{SSI}_3(\game)-\operatorname{SSI}_4(\game)=\frac{1}{2}.
$$
By an computation analogue to Subsection~\ref{subsec_SSI_median_2} we can confirm this value directly. For completeness we
give the entire calculation below.

\subsubsection{$\pi\in\Big\{(1,2,3,4),(1,2,4,3),(1,3,2,4),(1,3,4,2),(1,4,2,3),(1,4,3,2)\Big\}$:} For permutations $\pi$, where $\pi(1)=1$,
we have $\game\left(\overline{\tau}(x,\pi^{-1}(1)-1)\right)=1$, $\game\left(\overline{\tau}(x,\pi^{-1}(1))\right)=1$,
$\game\left(\underline{\tau}(x,\pi^{-1}(1))\right)=0$, and $\game\left(\underline{\tau}(x,\pi^{-1}(1)-1)\right)=0$, so that   
the value of the $4$-fold integral is given by $0$.

\subsubsection{$\pi\in\Big\{(2,1,3,4),(2,1,4,3),(3,1,2,4),(3,1,4,2),(4,1,2,3),(4,1,3,2)\Big\}$:} For permutation $\pi=(2,1,3,4)$ 
we consider the two cases $x_2\le x_1$ and $x_2\ge x_1$. In both cases we have $\game\left(\overline{\tau}(x,\pi^{-1}(1)-1)\right)=1$ 
and $\game\left(\underline{\tau}(x,\pi^{-1}(1)-1)\right)=0$. For $x_2\le x_1$ we have $\game\left(\overline{\tau}(x,\pi^{-1}(1))\right)=x_1$ 
and $\game\left(\underline{\tau}(x,\pi^{-1}(1))\right)=x_2$. For $x_2\ge x_1$ we have $\game\left(\overline{\tau}(x,\pi^{-1}(1))\right)=x_2$ 
and $\game\left(\underline{\tau}(x,\pi^{-1}(1))\right)=x_1$. Thus the corresponding $4$-fold integral is given by
\begin{eqnarray}
  \!\!\!\!\!\!\!&&\int_0^1\left(1-\left(\int_{0}^{x_1}x_1\,\operatorname{d}x_2+\int_{x_1}^{1}x_2\,\operatorname{d}x_2\right)
  +\left(\int_{0}^{x_1}x_2\,\operatorname{d}x_2+\int_{x_1}^{1}x_1\,\operatorname{d}x_2\right)-0\right)\operatorname{d}x_1\nonumber\\
  \!\!\!\!\!\!\!&&=  \int_0^1  \left(\frac{1}{2}+x_1(1-x_1)\right)\operatorname{d}x_1=\frac{2}{3}.
\end{eqnarray}
Due to symmetry we obtain the same values for all permutations $\pi$ with $\pi(2)=1$.

\subsubsection{$\pi\in\Big\{(2,3,1,4),(3,2,1,4),(2,4,1,3),(4,2,1,3),(3,4,1,2),(4,3,1,2)\Big\}$:} For permutation $\pi=(2,3,1,4)$ 
we consider the six strict orderings of $x_1,x_2,x_3$. In all six cases we have $\game\left(\overline{\tau}(x,\pi^{-1}(1)-1)\right)=1$ 
and $\game\left(\underline{\tau}(x,\pi^{-1}(1)-1)\right)=0$. If $x_1=\min(x_2,x_3)$ then 
$\game\left(\overline{\tau}(x,\pi^{-1}(1))\right)=\min(x_2,x_3)$ and $\game\left(\overline{\tau}(x,\pi^{-1}(1))\right)=x_1$ otherwise.
If $x_1=\max(x_2,x_3)$ then $\game\left(\underline{\tau}(x,\pi^{-1}(1))\right)=\max(x_2,x_3)$ and 
$\game\left(\underline{\tau}(x,\pi^{-1}(1))\right)=x_1$ otherwise.
Thus the corresponding $4$-fold integral is given by
\begin{eqnarray}
  \!\!\!\!\!\!\!&&\int_0^1\Big(1-\left(\int_{x_1}^1\int_{x_1}^1\!\!\min(x_2,x_3)\operatorname{d}x_2\,\operatorname{d}x_3+\int_{0}^{x_1}\int_{x_3}^{1}\!\!x_1\,\operatorname{d}x_2\,\operatorname{d}x_3+\int_{0}^{x_1}\int_{x_2}^{1}\!\!x_1\,\operatorname{d}x_3\,\operatorname{d}x_2 \right)\nonumber\\
  \!\!\!\!\!\!\!&&+\left(\int_0^{x_1}\int_0^{x_1}\!\!\max(x_2,x_3)\operatorname{d}x_2\,\operatorname{d}x_3 +\int_{x_1}^{1}\int_{0}^{x_3}\!\!x_1\,\operatorname{d}x_2\,\operatorname{d}x_3+\int_{x_1}^{1}\int_{0}^{x_2}\!\!x_1\,\operatorname{d}x_3\,\operatorname{d}x_2 \right)-0\Big)\operatorname{d}x_1\nonumber\\
  \!\!\!\!\!\!\!&&=\int_0^1 \left(\frac{2}{3}-x_1^2+x_1\right)\operatorname{d}x_1=\frac{5}{6}
\end{eqnarray}
Due to symmetry we obtain the same values for all permutations $\pi$ with $\pi(3)=1$.

\subsubsection{$\pi\in\Big\{(2,3,4,1),(3,2,4,1),(2,4,3,1),(4,2,3,1),(3,4,2,1),(4,3,2,1)\Big\}$:} For permutation $\pi=(2,3,4,1)$ we 
have $\game\left(\overline{\tau}(x,\pi^{-1}(1))\right)=\game\left(\underline{\tau}(x,\pi^{-1}(1))\right)$, 
$\game\left(\overline{\tau}(x,\pi^{-1}(1)-1)\right)=\max(x_2,x_3,x_4)$, and 
$\game\left(\underline{\tau}(x,\pi^{-1}(1)-1)\right)=\min(x_2,x_3,x_4)$.
Thus the corresponding $4$-fold integral is given by
\begin{equation}
  \int_0^1\left(\int_{x=(x_2,x_3,x_4)\in[0,1]^3}\!\!\!\!\!\!\!\!\!\!\!\!\!\!\!\!\!\!\!\!\!\!\!\!\left(\max(x_2,x_3,x_4)-\min(x_2,x_3,x_4)\right)\operatorname{d}x\right)\operatorname{d}x_1
  =\int_0^1\frac{\operatorname{d}x_1}{2}=\frac{1}{2}.
\end{equation}
Due to symmetry we obtain the same values for all permutations $\pi$ with $\pi(4)=1$.

\subsubsection{Summarizing the $24$ permutations for $\operatorname{SSI}_1(\game)$}
$$
  \operatorname{SSI}_1(\game)=\frac{1}{4!}\cdot\left(6\cdot 0+6\cdot \frac{2}{3}+6\cdot\frac{5}{6}+6\cdot\frac{1}{2}\right)=\frac{12}{24}=\frac{1}{2}.
$$

\section{Determining the $\operatorname{Nuc}$ for two continuous simple games}
\label{appendix_nucleolus}

From the binary case we can learn that very often it suffices to just minimize the maximum excess in order to compute
the nucleolus of a simple game. So for example $\hat{\game}$ we maximizing the excess
$$
  \frac{x_1^2+2x_2^2+3x_3^2}{6}-w_1x_1-w_2x_2-(1-w_1-w_2)x_3
$$
subject to the constraints $0\le x_1,x_2,x_3 \le 1$. With the aid of the corresponding Lagrange function
$L(x_1,x_2,x_3,\alpha_1,\alpha_2,\alpha_3,\beta_1,\beta_2,\beta_3)=$
\begin{eqnarray*}
  \!\!\!\!\!&&\frac{x_1^2+2x_2^2+3x_3^2}{6}-w_1x_1-w_2x_2-(1-w_1-w_2)x_3+\alpha_1x_1\\
  \!\!\!\!\!&&+\alpha_2x_2+\alpha_3x_3-\beta_1(x_1-1)-\beta_2(x_2-1)-\beta_3(x_3-1),
\end{eqnarray*}
we conclude that the maximum excess is attained at one of the solutions of the following equation system:
\begin{eqnarray*}
  \frac{\partial L}{\partial x_1} (x_1,\dots,\beta_3) =\frac{x_1}{3}-w_1+\alpha_1-\beta_1&\overset{!}{=}&0\\ 
  \frac{\partial L}{\partial x_2} (x_1,\dots,\beta_3) =\frac{2x_2}{3}-w_2+\alpha_2-\beta_2&\overset{!}{=}&0\\
  \frac{\partial L}{\partial x_3} (x_1,\dots,\beta_3) =x_3-w_3+\alpha_3-\beta_3&\overset{!}{=}&0\\
  \alpha_1x_1&\overset{!}{=}&0\\
  \alpha_2x_2&\overset{!}{=}&0\\
  \alpha_3x_3&\overset{!}{=}&0\\
  \beta_1(x_1-1)&\overset{!}{=}&0\\
  \beta_2(x_2-1)&\overset{!}{=}&0\\
  \beta_3(x_3-1)&\overset{!}{=}&0,
\end{eqnarray*}
where we have set $w_3=1-w_1-w_2$. The resulting $27$ solutions $(x_1,x_2,x_3)$ are given by
\begin{eqnarray*}
  x_1=3w_1 \quad\vee\quad x_1=0 \quad\vee\quad x_1=1  \\
  x_2=\frac{3w_2}{2} \quad\vee\quad x_2=0 \quad\vee\quad x_2=1 \\
  x_3=w_3 \quad\vee\quad x_3=0 \quad\vee\quad x_3=1
\end{eqnarray*}
For each of these $27$ combinations we can compute the excess $e^\game(x,w)$, where $x$ is specified by $w$. Now we are
just looking at the $8$ solutions where $x\in\{0,1\}^3$ and compute the respective excesses:
\begin{eqnarray*}
  e^\game\Big((0,0,0),w\Big) &=& 0,\\
  e^\game\Big((1,0,0),w\Big) &=& \frac{1}{6}-w_1,\\
  e^\game\Big((0,1,0),w\Big) &=& \frac{1}{3}-w_2,\\
  e^\game\Big((0,0,1),w\Big) &=& \frac{1}{2}-w_3,\\
  e^\game\Big((1,1,0),w\Big) &=& w_3-\frac{1}{2},\\
  e^\game\Big((1,0,1),w\Big) &=& w_2-\frac{1}{1},\\
  e^\game\Big((0,1,1),w\Big) &=& w_1-\frac{1}{6},\\
  e^\game\Big((1,1,1),w\Big) &=& 0.   
\end{eqnarray*}  
Thus we have that the maximum excess is at least
$$
  \max\left(\left|\frac{1}{6}-w_1\right|,\left|\frac{1}{3}-w_2\right|,\left|\frac{1}{2}-w_3\right|\right)\ge 0.
$$
More precisely, each choice of $w\neq \frac{1}{6}\cdot (1,2,3)$ leads to a maximum excess larger than $0$. Next we show that
choosing $w=\frac{1}{6}\cdot (1,2,3)$ yields a maximum excess of $0$, so that
$\operatorname{Nuc}(\game)=\left\{\frac{1}{6}\cdot (1,2,3)\right\}$. It remains to prove
$e^\game\Big(x,\frac{1}{6}\cdot(1,2,3)\Big)\le 0$ for all $x\in[0,1]^3$. We may check all $27$ solutions and verify that
the excess at these points is at most $0$ or simply rewrite the formula for the maximum excess at 
$w=\frac{1}{6}\cdot(1,2,3)$ to
$$
  e^\game\Big(x,\frac{1}{6}\cdot(1,2,3)\Big)=-\frac{x_1(1-x_1)}{6}-\frac{x_2(1-x_2)}{3}-\frac{x_3(1-x_3)}{2},  
$$
which is obviously upper bounded by zero for all $x\in[0,1]^3$.

\medskip

For our second example minimizing the maximum excess does not help very much. Using the fact that the geometric mean
is always at most as large as the arithmetic mean we deduce
$$
  x_1x_2^2x_3^3\le\sqrt[n]{x_1x_2^2x_3^3}\le\frac{x_1+2x_2+3x_3}{n}
$$
for all integers $n$. If $w_1,w_2,w_3>0$ the right hand side is at most $w_1x_1+w_2x_2+w_3x_3$ for suitably large $n$.
Using an easy but elaborated argument for the cases where some of the $w_i$ but not all are zero, we can conclude
$e^\game(x,w)\le 0$ for all $x\in[0,1]^3$ and all $w\in[0,1]^3$ satisfying $\Vert w\Vert_1=1$. Thus from the maximum excess
we can only conclude the trivial implication $\operatorname{Nuc}(\game)\in\left\{w\in[0,1]^3\,:\,\Vert w\Vert_1\right\}$.

In order to get an idea of how the nucleolus may be computed in similar cases, we have considered the more simple 
two voter example $\game(x_1,x_2)=x_1x_2^2$. Here minimizing the maximum excess is of little use to. A feasible approach
might be to compute an exact expression for the excess function in the domain $[c,1]$, where $c<1$ is near to one, 
parametric in $w_1$ and $w_2=1-w_1$. Using a somewhat elaborated case distinction and assuming a $c$ sufficiently close to $1$,
we were able to state an exact formula for the excess function using a sum of several integrals. Those integrals 
could be evaluated numerically for fix values of $w_1,w_2$. By plotting and comparing the excess functions for certain
weights, we were able to compute the numerical bounds $0.4553 \le y_1\le 0.4555$ and $0.5545 \le y_2\le 0.5547$ for
all $(y_1,y_2)\in\operatorname{Nuc}(\game)$.

\end{document}